\documentclass[twocolumn, pra]{revtex4-1}
\usepackage{graphicx}
\usepackage{amssymb}
\usepackage{epstopdf}
\usepackage{amsmath}
\usepackage{color}
\makeatletter

\@ifundefined{textcolor}{}
{
 \definecolor{BLACK}{gray}{0}
 \definecolor{WHITE}{gray}{1}
 \definecolor{RED}{rgb}{1,0,0}
 \definecolor{GREEN}{rgb}{0,1,0}
 \definecolor{BLUE}{rgb}{0,0,1}
 \definecolor{CYAN}{cmyk}{1,0,0,0}
 \definecolor{MAGENTA}{cmyk}{0,1,0,0}
 \definecolor{YELLOW}{cmyk}{0,0,1,0}
}

\usepackage[colorlinks,
            linkcolor=black,
            anchorcolor=black,
            citecolor=black
            ]{hyperref}
\usepackage{ulem}           
            
\makeatother

\begin{document}

\title{Engineering Quantum Spin Liquids and Many-Body Majorana States with a Driven Superconducting Box Circuit}

\author{Fan Yang$^1$, Lo\" ic Henriet$^2$, Ariane Soret$^{1,3}$, Karyn Le Hur$^1$}
\affiliation{$^1$ CPHT, Ecole Polytechnique, CNRS, Universit\' e Paris-Saclay, Route de Saclay, 91128 Palaiseau, France}
\affiliation{$^2$ ICFO-Institut de Ciencies Fotoniques, The Barcelona Institute of Science and Technology, 08860 Castelldefels (Barcelona), Spain}
\affiliation{$^3$ Department of Physics, Technion Israel Institute of Technology, 32000 Haifa, Israel}

\date{\today}

\begin{abstract}
We design a driven superconducting box with four spins S=1/2 (qubits) such that coupled devices can give insight on the occurrence of quantum spin liquids and many-body Majorana states. Within one box or island, we introduce a generalized nuclear magnetic resonance algorithm to realize our models and study numerically the spin observables in time as well as the emergent gauge fields. We discuss the stability of the box towards various detuning effects and we include dissipation effects through a Lindblad master equation. Coupling boxes allows us to realize quantum spin liquid phases of Kitaev ${\cal Z}_2$ spin models  in various geometries with applications in the toric code. Quantum phase transitions and Majorana physics might be detected by measuring local susceptibilities. We show how to produce a N\' eel state of fluxes by coupling boxes and we address the role of local impurity fluxes leading to random Ising models. We also present an implementation of the Sachdev-Ye-Kitaev Majorana model in coupled ladder systems.
\end{abstract}

\maketitle

\section{Introduction}

Majorana fermions have revived attention due to possible applications in quantum information as protected qubits \cite{Franz,Beenakker,Marcus,Bert,Matthew,Yazdani,review} and surface codes with ${\cal Z}_2$ variables \cite{AltlandEgger,Terhal,Fu}. We design a Majorana box starting from a superconducting four-site circuit \cite{SPEC,Yale,Roushan} with the goal to engineer quantum spin liquids and many-body Majorana states encoded in spin-1/2 degrees of freedom. Starting with four transmon qubits, we present a Nuclear Magnetic Resonance (NMR) double-period protocol to realize the box.  We study the quantum dynamics in time to implement the required protocols and to detect the ${\cal Z}_2$ gauge fields through spin variables. A system of three transmons in cQED has been realized recently \cite{Roushan}, with possible applications in topological phases \cite{Koch,Review2017}.

These boxes could be used in variable geometries from quantum impurity systems to tunable ladder and plaquette models. Ensembles of square-plaquette models have been realized in ultra-cold atoms \cite{Munich} to emulate an Anderson Resonating Valence Bond spin-liquid state \cite{Anderson}, and have been shown theoretically to be related to {$d$}-wave superconductivity (superfluidity) in the Hubbard model close to the Mott state \cite{KarynMaurice}.  The design of such Majorana boxes addresses challenging questions regarding the choice of couplings. Experiments in superconducting circuit quantum electrodynamics (QED) architectures \cite{zhong} and in ultra-cold atoms \cite{Dai} report progress in engineering four-body interactions inspired by theoretical efforts \cite{Fisher,Zurich}. Engineering four-body interactions is also at the heart of our proposal to realize gauge fluxes, loop currents, and Majorana states in quantum spin liquids. 
 
\begin{figure}[t]
  \begin{center}
    \includegraphics[width=8.5cm]{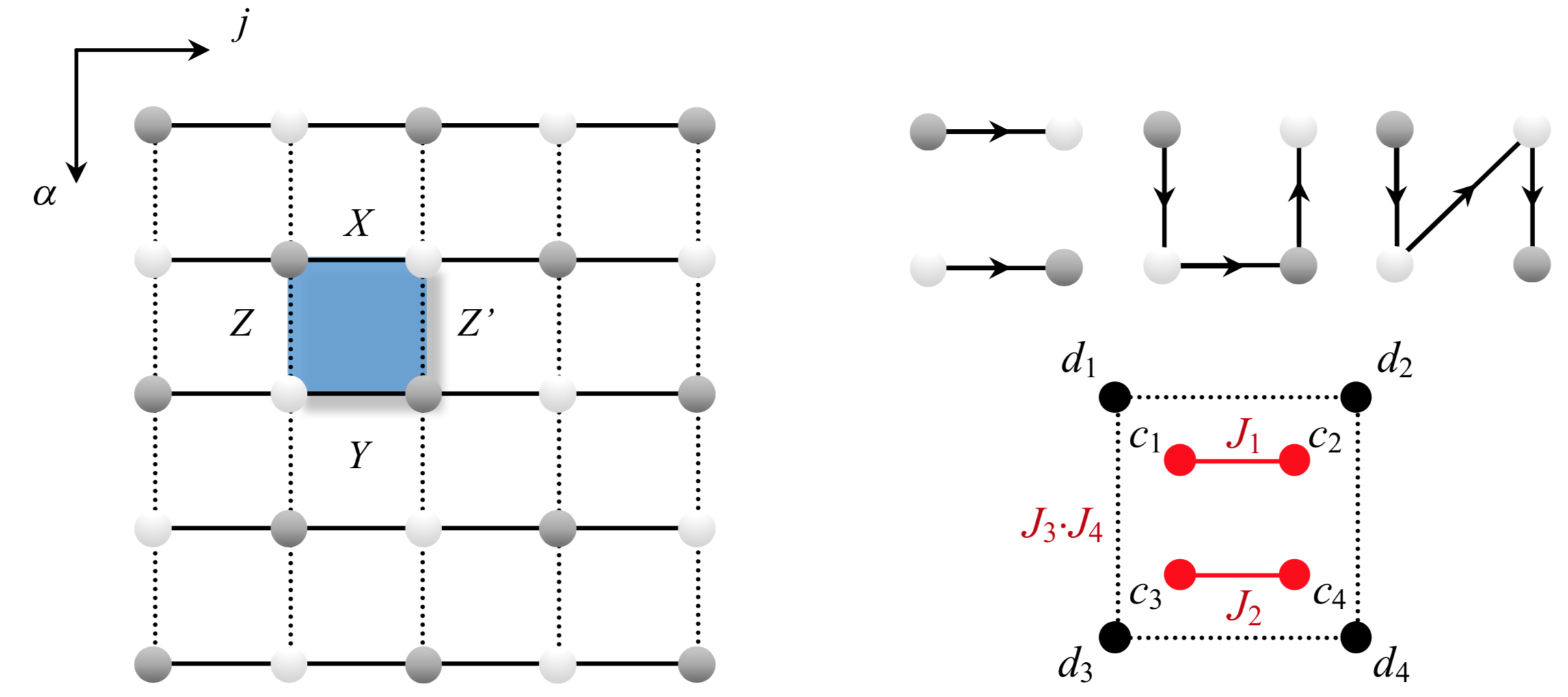}
          \end{center}
  \vskip -0.5cm \protect\caption[]
  {(color online) (Left) Two-dimensional lattice built from coupled boxes with $\mathcal{Z}_2$ symmetry: $XYXY$ alternating Ising couplings along horizontal bonds and $ZZ'ZZ'$ couplings on vertical bonds. (Right top) Different configurations of Jordan-Wigner strings for one unit cell. (Right bottom) Majorana representation: $J_1$, $J_2$, $J_3 (J_4) $ denote respectively the $X$, $Y$ and $Z$ coupling constants. When $|J_1|, |J_2| \gg |J_3|, |J_4|$, $c$ Majorana particles are gapped at high energies and the $d$ Majorana fermions describe the state of gauge fields in each unit cell or square plaquette.}
    \label{fig:lattice}
    \vskip -0.5cm
\end{figure}

Within our framework, a lattice system can be built by coupling a number of boxes, forming then  coupled-ladder models as in Fig.~\ref{fig:lattice}. Coupled boxes could allow us to re-build the Kitaev ${\cal Z}_2$ quantum spin model of the honeycomb lattice \cite{Kitaev} in ladder systems \cite{KFA,Feng,Basel,Smitha,Yao,Motrunich} with potential applications in the toric code \cite{Kitaevcode} and other surface codes \cite{Martinis}. These models have stimulated the discovery of {quantum} materials \cite{Dima,Klanjsek,Jackeli,Simon,Mendels,Cava} as well as the design of ultra-cold {atoms} \cite{Duan,Zoller} and other superconducting architectures \cite{Doucot,Neel,Barends}.  It is important to mention other proposals of Majorana boxes related to topological superconducting wires \cite{AltlandEgger,Terhal} and topological superconductors \cite{Fu}.  Realizing a pure four-body Majorana fermion coupling also allows us to emulate the Sachdev-Ye-Kitaev (SYK) model \cite{SachdevYe,OlivierAntoine,KitaevnewJosephine} with coupled boxes as elaborated below. The SYK model, which involves a (long-range and disordered) coupling between four Majorana fermions, has attracted attention theoretically in high-energy \cite{Polchinski,Maldacena,Witten} and low-energy physics \cite{Pikulin,Alicea,Laflamme} due to possible black-hole gravity holographic correspondence \cite{KitaevnewJosephine} and link to quantum chaos \cite{Stanford}.  Only a few realizations of the SYK Majorana model have been discussed so far \cite{Pikulin,Alicea,Laflamme}. SYK spin models could also bring light on quantum glasses \cite{OlivierAntoine}. 

Before proceeding to the engineering side of the circuit network, it is relevant to introduce the mapping of $\mathcal{Z}_2$ (or Ising like) spin models to Majorana fermions and the notion of flux states.  On horizontal bonds, as shown in Fig. 1, there are $XYXY$ alternating Ising interactions with coupling constants $J_1$ and $J_2$.  For the vertical bonds, we allow $ZZ'ZZ'$ couplings with strengths $J_3$ and $J_4$. A unit cell of four sites is depicted as the blue box.  A general lattice of Fig.~\ref{fig:lattice} holds a class of exactly solvable models for  quantum spin liquids. By setting $Z' = 0$, the brick-wall lattice recovers the Kitaev honeycomb model. Multi-leg ladders can then be addressed, as well as the passage from one to two dimensions, or higher-dimensional lattices. 

The sites are labelled through the $j$-th column and $\alpha$-th row, forming two sublattices $A$ ($j+\alpha  = \text{even}$) and $B$ ($j+\alpha  = \text{odd}$). We can perform the Jordan-Wigner transform, $\sigma_j^\dagger = a_j^\dagger e^{i \pi \sum_{l<j} a_l^\dagger a_l}, \sigma_j^- = a_j e^{i \pi \sum_{l<j} a_l^\dagger a_l}$. The ground state, by analogy {with} a particle in a box in quantum mechanics, shows no excitation along the string \cite{KFA,Feng}. Each spin is represented by a fermion operator and therefore $a^{\dagger}_l a_l$ can take values 0 or 1: eigenvalues for $\sigma_j^z=2a^{\dagger}_j a_j-1$ are $\pm 1$. Each fermion can be seen as two Majorana fermions $c_j$ and $d_j$:
\begin{gather} j \in A
\begin{cases}
c_j = i(a_j^\dagger - a_j) \\
d_j = a_j^\dagger + a_j
\end{cases}
\hskip -0.3cm
{; j} \in B 
\begin{cases}
c_j = a_j^\dagger + a_j\\
d_j = i(a_j^\dagger - a_j).
\end{cases}
\label{eq:maj}
\end{gather}
In a square of four sites, we obtain
\begin{gather}
\mathcal{H}_K = J_1\sigma_1^x\sigma_2^x + J_2\sigma_3^y\sigma_4^y + J_3 \sigma_1^z\sigma_3^z + J_4 \sigma_2^z\sigma_4^z  \notag \\
= -i J_1 c_1c_2 + iJ_2 c_3c_4 - i J_3D_{1,3}c_1c_3-iJ_4D_{2,4}c_2c_4
\end{gather}
with $D_{1,3} = -i d_1d_3$ and $D_{2,4} = -i d_2d_4$. The couplings $J_1$ and $J_2$ are ferromagnetic (or $J_1,J_2<0$), and the couplings $J_3$ and $J_4$ are adjustable couplings through the fluxes $\Phi_3$ and $\Phi_4$ in Fig. 2.
Different string paths in Fig.~\ref{fig:lattice} (Right top) give identical results. This result has been confirmed rigorously for the ladder geometries \cite{KFA}. It is relevant to note that the $d$-Majorana fermions enter through the emergence of
$\mathcal{Z}_2$ gauge fields: $D_{1,3}$ and $D_{2,4}$ commute with $\mathcal{H}_{K}$ and take values $\pm 1$. On a square unit cell,
then we can define the associated flux operator 
\begin{equation}
\mathcal{P}_d = d_1d_2d_3d_4 = D_{1,3}D_{2,4}.
\end{equation} 
This flux operator acting on a unit square cell, and encoded with the {$d$-}Majorana ${\cal Z}_2$ variables, in our representation intervenes through the product of parity operators of two d-Majorana fermions forming the vertical bonds. 

The limit of weak vertical bonds $|J_1|, |J_2| \gg |J_3|, |J_4|$  (see Fig.~\ref{fig:lattice} Right bottom) is of particular interest to us.  The {$c$-}Majorana fermions are gapped describing the formation of valence bonds in the spin language between sites 1 and 2, and 3 and 4, respectively.  In addition, $-i c_1 c_2=+1$ and $i c_3 c_4=+1$ such that we can define the operator $\mathcal{P}_c=c_1 c_2 c_3 c_4=+1$.  The $d$-Majorana particles will be coupled in a 4-body coupling, as in the SYK model. More precisely, the leading-order term in the perturbation theory gives $- {J_3J_4}/{(|J_1| + |J_2|)} \sigma_1^z \sigma_2^z \sigma_3^z \sigma_4^z =- {J_3J_4}/{(|J_1| + |J_2|)}\mathcal{P}_d \mathcal{P}_c$ with $\mathcal{P}_c=1$. If $J_3J_4 > 0, \mathcal{P}_d = 1$ corresponds to the $\pi$-flux configuration in a square unit cell, in agreement with the Lieb's theorem \cite{Lieb}; otherwise $\mathcal{P}_d = -1$ relates to the $0$ flux.

Below, we show how to detect the gauge fields, at the level of one box and a few boxes.  It is also relevant to note that by assembling boxes, one can then build a spin model, which turns {out to be} a quantum spin liquid with a {$\pi$-}flux ground state.  A staggered flux order has also been suggested for high-{$T_\text{c}$} cuprates  \cite{AffleckMarston}.  Recent efforts in quantum materials report the observation of orbital loop currents in Mott materials with spin-orbit coupling \cite{Philippe} by analogy {with} cuprates \cite{cuprates}. Here, we can tune parameters in the spin system and adjust the ground state to have such a $\pi$ flux. The coupled-ladder geometry then presents some tunability.

The paper is organized as follows. In Sec.~\ref{sec:alg}, we show how to engineer $\mathcal{H}_K$ with superconducting circuits and introduce our main algorithm. In Sec. III, we perform numerical tests on the time-dependent Hamiltonian, and study stability of the box towards detuning and dissipation effects. Then, we address measurements of gauge fields through spin degrees of freedom.  Disorder (local impurities) in the gauge fields can be implemented through magnetic fluxes
and through time-dependent protocols.  In Sec. IV, we discuss applications for an ensemble of coupled boxes, such as the realization of Kitaev spin models and the emergence of N\' eel (Ising-like) order for the gauge fields. We also address relations with Wen's toric code \cite{Wen} and possible SYK loop models. In Sec. V, we briefly summarize our results and appendices are devoted
for additional technical calculations and summary tables. 

\section{Algorithm on an island}
\label{sec:alg}
\subsection{Physics of a box}

\begin{figure}[t]
  \begin{center}
    \includegraphics[width=5cm]{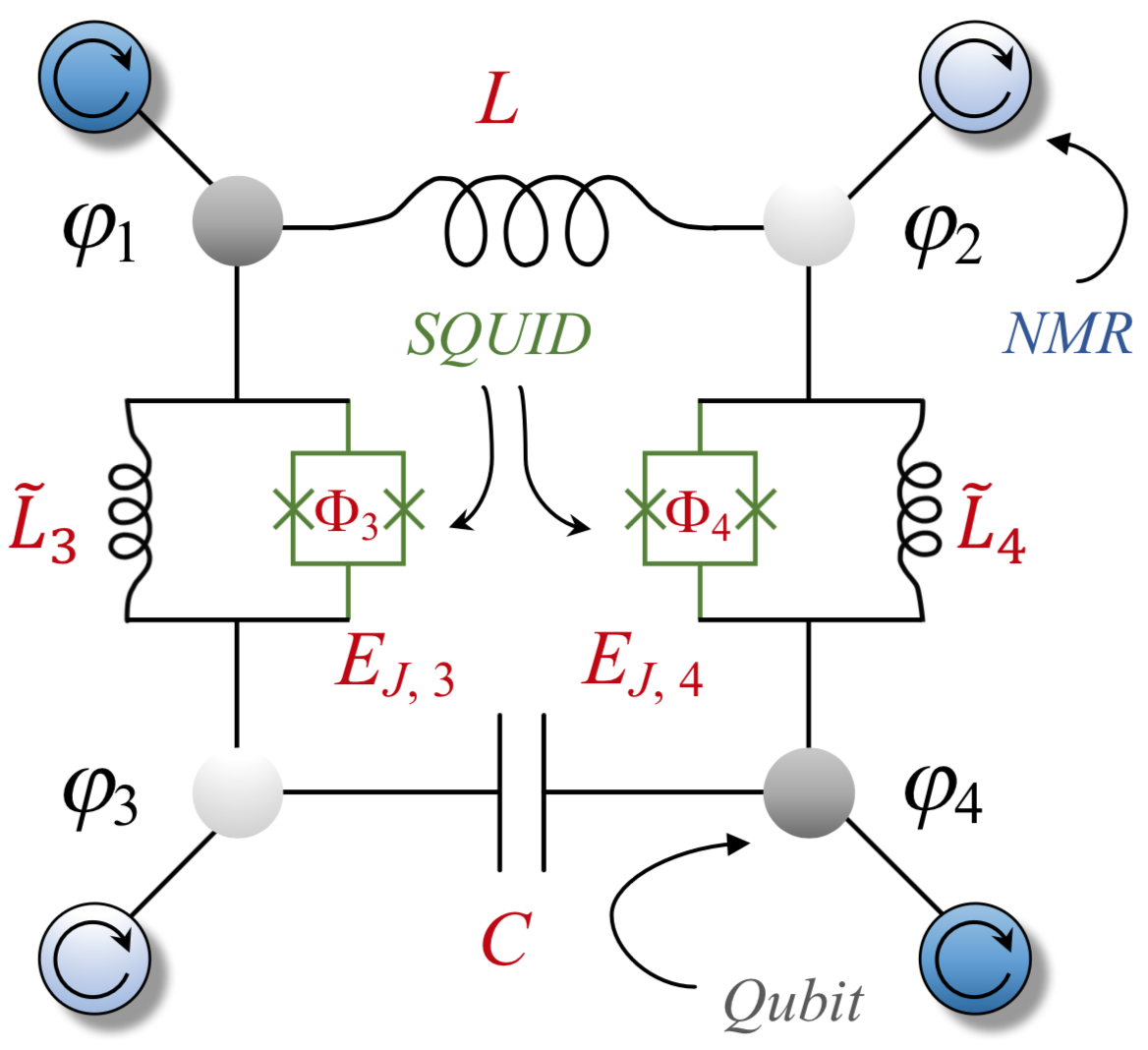}
     \includegraphics[width=7.5cm]{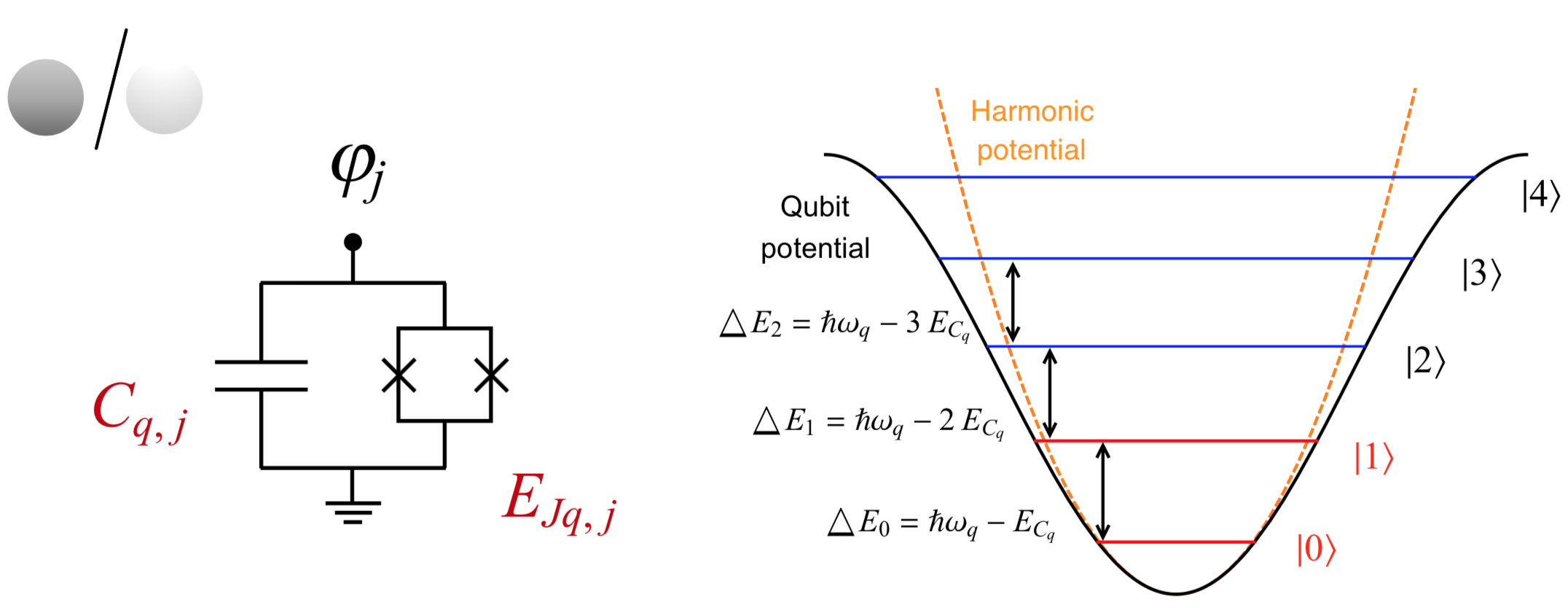}
       \includegraphics[width=7.5cm]{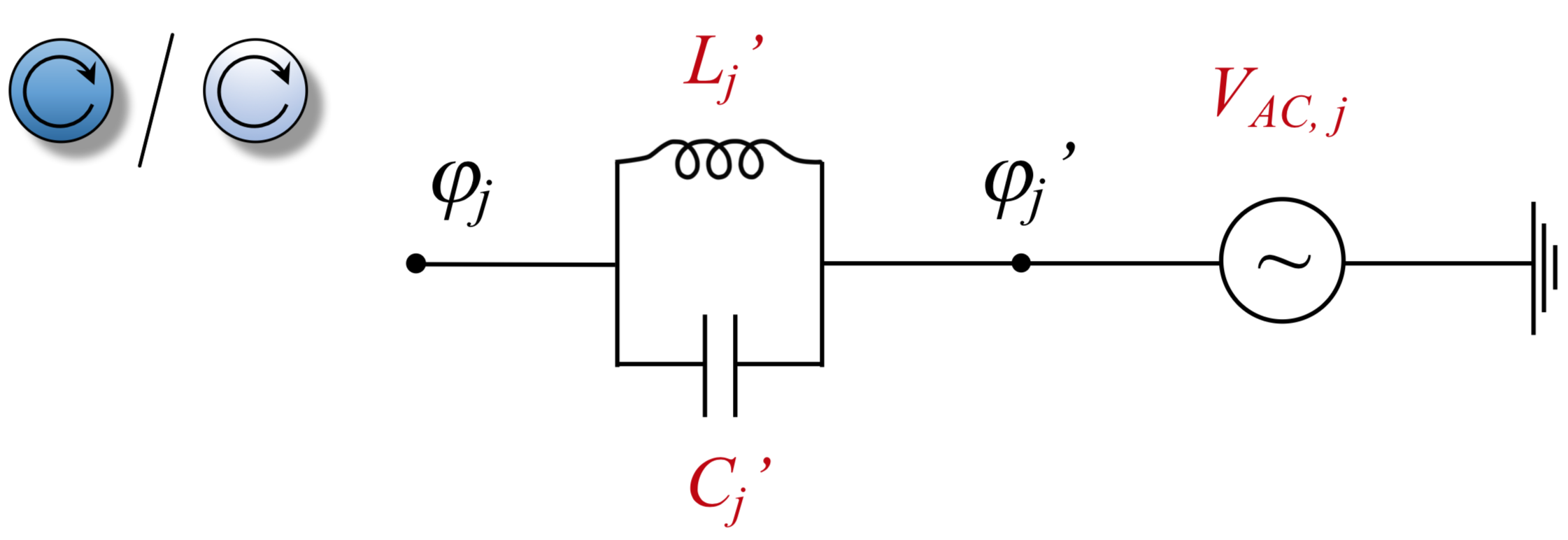}
       \end{center}
  \vskip -0.5cm \protect\caption[]
  {(color online) (Top) We engineer  $X$ and $Y$ Ising couplings through inductance $L$ and capacitance $C$ on horizontal bonds, $Z$ couplings with SQUIDs and auxiliary inductances $\tilde{L}$ on vertical bonds;  (Middle left) Structure of on-site transmon qubits: composed of two Josephson junctions and a capacitance in parallel; (Middle right) {Spectrum} of transmon qubits realized with the two lowest levels; (Bottom) Structure of the generalized NMR device: producing a circularly polarized driven field. Different colors of qubits (grey and white) and NMR fields (dark blue and light blue) indicate two distinct sets of frequency patterns for sublattices $A$ and $B$.}
    \label{fig:device}
    \vskip -0.5cm
\end{figure}

First, we introduce the physical structure of one box in Fig.~\ref{fig:device}. Within a cell of four sites, we denote the superconducting phases as $\hat{\varphi}_j \ (j = 1,4 \in \{A\}; j = 2,3 \in \{B\})$. One box  can be decomposed into three parts: the on-site transmon, the local NMR device and the inter-site couplings.  Fig.~\ref{fig:device} Middle shows the internal structure of each site. We build a transmon qubit on the site $j$ via sets of capacitances and Josephson junctions
$\{C_{q, A}, E_{J_{q, A}}\}$ and $\{C_{q,B}, E_{J_{q,B}}\}$, of which the resonance (plasma) frequencies will be adjusted accordingly. The qubit Hamiltonian reads:
\begin{gather}
{\cal H}_{q,j} = \frac{C_{q,j}\phi_0^2}{2} \dot{\hat{\varphi}}_j^2 - E_{J_q,j} \cos \hat{\varphi}_j \label{eq:hq},
\end{gather}
where $\phi_0= \hbar/(2e)$ denotes the rescaled quantum of flux and $E_{J_q,j}$ represents the Josephson energy of the internal junction.

In Fig.~\ref{fig:device} Bottom, we then connect each node $j$ to an inductance $L_j'$ and a capacitance $C_j'$  followed by an AC source of voltage, generating a time-dependent NMR field
\begin{gather}
{\cal H}_{\text{NMR},j} = E_{L',j} \left( {\varphi}'_j  - \hat{\varphi}_j \right)^2 +  \frac{C'\phi_0^2}{2} \left( \dot{{\varphi}}'_j  - \dot{\hat{\varphi}}_j \right)^2 + E_{V_{AC},j}.
\label{eq:hnmr}
\end{gather}
The main purpose of this field is to cancel the local magnetic field in the rotating frame, as we will show later.  The time dependence of ${\cal H}_{\text{NMR},j}$ is encoded in parameters ${\varphi}'_j$ and ${\dot{\varphi}}'_j$ which 
satisfy the relations: $\phi_0\dot{{\varphi}}'_j = -V_{AC,j} = -V_{0,j} \sin \left(\omega_j t\right), {\varphi}'_j = \int  dt \ \dot{{\varphi}}'_j = {V_{0,j}} \cos \left(\omega_j t\right)/({\phi_0\omega_j})$. We choose to apply this NMR device because it preserves the
$\mathcal{Z}_2$ symmetry of the Hamiltonian.  This protocol is then distinct {from} the protocol used in Ref. \cite{Roushan} for the 3-qubit system. 

For the interaction part, as can be seen from Fig.~\ref{fig:device} Top, horizontal bonds of the box are coupled by an inductance $L$ and a capacitance $C$ to engineer respectively $X$ and $Y$ couplings. The corresponding interaction Hamiltonians take the form
\begin{gather}
{\cal H}_{L} = E_L \left( \hat{\varphi}_{2} - \hat{\varphi}_{1}\right)^2, \quad {\cal H}_{C} = \frac{C\phi_0^2}{2} \left( \dot{\hat{\varphi}}_{4} - \dot{\hat{\varphi}}_{3}\right)^2
 \end{gather}
 with $E_L = {\phi_0^2}/{(2L)}$. 
 
Realizing pure $Z$ couplings on vertical bonds can be achieved through SQUIDs. The SQUIDs (with characteristic Josephson energies $E_{J,3}$ and $E_{J,4}$) are controlled {via} applied magnetic fields $\Phi_3$ and $\Phi_4$, and {we add} auxiliary inductances $\tilde{L}_3$  and $\tilde{L}_4$ to compensate the additional $X$ couplings (see Fig. 2). For instance, on the vertical bond $(1,3)$, the interaction energy of the SQUID has the form
\begin{gather}
{\cal H}_{S,3} = - E_{J,3} \cos\left(\hat{\varphi}_1 - \hat{\varphi}_3 \right),
\label{eq:hs}
\end{gather}
while the auxiliary inductance $\tilde{L}_3$ contributes to
\begin{gather}
{\cal H}_{\tilde{L},3} = E_{\tilde{L}} \left( \hat{\varphi}_{1} - \hat{\varphi}_{3}\right)^2,
\end{gather}
 with $E_{\tilde{L}} = {\phi_0^2}/{(2\tilde{L})}$.
We study perturbations arising from vertical bonds in Sec.~\ref{sec:per}.

The total Hamiltonian can now be written as
\begin{gather}
{\cal H} = \sum_{j=1}^4 {\cal H}_{q,j} + {\cal H}_{{\text{NMR},j}} + {\cal H}_L + {\cal H}_C + {\cal H}_{S} + {\cal H}_{\tilde{L}}.
\end{gather}

\subsection{Quantized Hamiltonian}

We start from the quantization \cite{Yale} of the transmon qubit Hamiltonian $\mathcal{H}_{q,j}$, which behaves as harmonic oscillators with anharmonicity from Josephson junctions. Expanding the nonlinear cosine potential in Eq.~(\ref{eq:hq}) to the fourth order and choosing the bosonic representation: $[\hat{\varphi}_j, \hat{\pi}_l] = i\hbar \delta_{j,l},  \hat{\varphi}_j =  ( b_j^\dagger + b_j )/{\lambda_j},\hat{\dot{\varphi}}_j = ( b_j^\dagger - b_j)({-e\lambda_j})/(i\phi_0C_{q,j})$ with conjugate momentum $\hat{\pi}_j = {\phi_0^2} C_{q,j}\dot{\hat{\varphi}}_j$, we reach
\begin{gather}
	 {\cal H}_{q,j }= - {E}_{J_{q,j}} + \hbar \omega_{q,j} \left( b_j^\dagger b_j + \frac{1}{2} \right) - \frac{E_{C_{q,j}}}{12} \left( b_j^\dagger + b_j\right)^4.
	 \label{eq:h4}
  \end{gather}
Here  we assume the system in the large $\lambda_j = ( E_{J_{q,j}}/(2E_{C_{q,j}}) )^{1/4} $ limit. $E_{C_{q,j}} = e^2/(2C_{{q,j}})$ depicts the charging energy associated with the transfer of a single electron. 
$\omega_{q,j} = \sqrt{8E_{C_{q,j}}E_{J_{q,j}}}/ \hbar$ is known as the Josephson plasma frequency ($\sim$ GHz corresponding to  $T \sim 0.1 \ \text{K}$). 
 
As shown in Fig.~\ref{fig:device} Middle right, we denote the eigenstates of a pure harmonic oscillator as $\left| n_j \right>$. Taking into account the leading-order correction from the quartic term in Eq.~(\ref{eq:h4}), the spectrum of a transmon is modified into $E_{n,j}  = -E_{{J_q,j}} + \hbar \omega_{q,j} \left( n_j + 1/2 \right) - E_{C_{q,j}} \left( 6n_j^2 + 6n_j + 3\right)/12$. The gap is decreasing between two successive energy levels: $\Delta E_{n,j} = E_{n+1,j} - E_{n,j}  = \hbar \omega_{q,j} - E_{C_q,j} \left( n_j+1\right)$. If we restrict the state of each transmon $j$ to the two lowest energy levels $\left| 0 \right>_j$ the quantum vacuum and $\left| 1 \right>_j$ the state with one quantum, a qubit will be formed. As transitions to higher levels are forbidden, $b_j$ become hard-core bosons obeying  $b_j^n = ( b_j^\dagger)^n = 0$ for any $n \ge 2$. It allows for a mapping to the spin-1/2 states for an individual site: $\left| 0 \right>_j \leftrightarrow \left| \downarrow \right>_j, \left| 1 \right>_j \leftrightarrow \left| \uparrow \right>_j, b_j^\dagger \leftrightarrow \sigma_j^+, b_j \leftrightarrow \sigma_j^-$ with $\left| \downarrow \right>_j$ and $\left| \uparrow \right>_j$ polarized along $z$ direction. In the spin space,
 \begin{gather}
\sigma_j^x = b_j^\dagger + b_j, \quad \sigma_j^y = \frac{1}{i} (b_j^\dagger - b_j), \quad \sigma_j^z = 2b_j^\dagger b_j -1.
\end{gather}
Eigenvalues of $\sigma_j^z$ are well fixed to $\pm 1$ since we restrict ourselves to the subspace where $b^{\dagger} b=0$ or $1$. 
Now,  the effective Hamiltonian of a transmon qubit acts as a strong local magnetic field
 \begin{gather}
{\cal H}_{q,j} \simeq\Delta E_{0,j} b_j^\dagger b_j = \epsilon_{q,j} \sigma_j^z, \label{eq:hqu}
 \end{gather}
 where $\epsilon_{q,j} = \Delta E_{0,j}/2 = (\hbar\omega_{q,j} - E_{C_{q,j}})/2$ characterizes the transition energy from $\left| 0 \right>_j$ to $\left| 1 \right>_j$. In the absence of {an} AC driving source, the spin system would be polarized meaning that all the transmon systems would be in the quantum vacuum. 

Through this quantization procedure, the NMR field is transformed into
 \begin{gather}
 \begin{split}
{\cal H}_{\text{NMR},j} =& - \frac{\hbar \omega_{L',j}}{2}\cos ( \omega_j t) \sigma_j^x -\frac{\hbar \omega_{C',j}}{2} \sin ( \omega_j t ) \sigma_j^y\\
&+ \left( \epsilon_{L',j} +  \epsilon_{C',j}\right) \sigma_j^z,
\end{split}
 \end{gather}
with the fast-oscillating terms $E_{L',j} ({\varphi}'_j)^2$, ${C'} (\phi_0 \dot{{\varphi}}'_j )^2/2$ and $E_{V_{AC},j}$ dropped out. For simplicity, all coefficients are listed in Appendix A. 
Furthermore,  we impose 
 \begin{gather}
 \omega_{L',j} = \omega_{C',j} = \omega_{1,j}
 \end{gather}
 to generate a circularly polarized field. [The stability in the presence of a small detuning from this condition is related to the discussion in Eq. (31).]
 
On the horizontal bonds, the interaction Hamiltonians become
\begin{gather}
{\cal H}_{L} = \epsilon_{L, A}  \sigma_{1}^z + \epsilon_{L, B}  \sigma_{2}^z + J_1 \sigma_{1}^x\sigma_{2}^x,\notag \\
{\cal H}_{C} = \epsilon_{C, B} \sigma_3^z + \epsilon_{C, A} \sigma_4^z + J_2 \sigma_3^y\sigma_4^y,
\end{gather}
where $J_1 < 0$ and $J_2 < 0$.

A more detailed analysis is needed for the vertical bonds. In the large $\lambda_j$ limit, $\hat{\varphi}_j$ can be viewed as a small quantum variable. We are allowed to ignore higher order contributions of the cosine potential in Eq.~(\ref{eq:hs}). To the fourth order, ${\cal H}_{S,3} = - E_{J,3} ( 1-  (\hat{\varphi}_1 - \hat{\varphi}_3 )^2/2! +  (\hat{\varphi}_1 - \hat{\varphi}_3 )^4/4!+ \cdots)$. The quadratic terms give arise to an effective $X$ coupling $\hat{\varphi}_1 \hat{\varphi}_3 \sim \sigma_1^x \sigma_3^x$ and a magnetic field $\hat{\varphi}_1^2 \sim \sigma_1^z, \hat{\varphi}_3^2 \sim \sigma_3^z$. For the quartic contribution, the only effective term  $\hat{\varphi}_1^2 \hat{\varphi}_3^2$ produces a $Z$ coupling $\sigma_1^z \sigma_3^z$. Thus,
\begin{gather}
{\cal H}_{S,3} = J_3 \sigma_1^z \sigma_3^z  + J_3^x \sigma_1^x \sigma_3^x + \epsilon_{J,1}\sigma_{1}^z + \epsilon_{J,3} \sigma_{3}^z,
\end{gather}
where $J_{3},J_3^x \propto -E_{J,3}$.  Both the signs and amplitudes of vertical couplings can be adjusted by the flux $\Phi_3$ inside the SQUID as $E_{J,3} \sim \cos ( {\Phi_{3}}/{(2\phi_0}))$.

At the same time, the auxiliary inductance $\tilde{L}_3$ gives a negative $X$ coupling
\begin{gather}
{\cal H}_{\tilde{L},3} = \tilde{J}^x_3 \sigma_1^x \sigma_3^x + \epsilon_{\tilde{L},A}\sigma_1^z +  \epsilon_{\tilde{L},B}\sigma_3^z.
\end{gather}
We can then reduce the vertical $X$ couplings to zero: 
\begin{gather}
\label{eq:xx}
J_3^x + \tilde{J}_3^x = 0,
\end{gather}
with the phase $\Phi_3/(2\phi_0) \in \left[\pi/2+2n\pi, 3\pi/2+2n\pi \right[, n \in \mathbb{Z}$ for a positive {${J}_3^x$}. It is the same case with bond $(2,4)$. 

 Combined with the local $\sigma_j^z$ field of the transmon qubit, the total effective Hamiltonian of the box becomes
  \begin{gather}
 	 {\cal H} = {\cal H}_K + {\cal H}_C(t),   \label{eq:hf}
	  \\ \notag {\cal H}_K = J_1^x \sigma_1^x\sigma_2^x + J_2\sigma_3^y\sigma_4^y + J_3\sigma_1^z\sigma_3^z + J_4\sigma_2^z\sigma_4^z,\notag \\ \notag
 	{\cal H}_C(t) = \sum_j \frac{\hbar \omega_{0,j}}{2} \sigma_j^z  -\frac{\hbar \omega_{1,j}}{2} \left(  \cos \left( \omega_j t \right) \sigma_j^x + \sin \left( \omega_j t \right) \sigma_j^y \right). 
  \end{gather}
The time-dependent Hamiltonian ${\cal H}_C(t)$ here is distinct from the capacitive Hamiltonian ${\cal H}_C$ introduced above in the intermediate steps of the reasoning. Generally, ${\hbar \omega_{0,j}}/{2} {= \epsilon_j} = \epsilon_{q,j} +\epsilon_{L',j} + \epsilon_{C',j} + \epsilon_{L,j} + \epsilon_{C,j} + \epsilon_{J,j} + \epsilon_{\tilde{L},j}$. 
The main contribution to $\omega_{0,j}$ arises from the qubit transition energy $\epsilon_{q,j}$. Other minor terms may vary depending on the geometries (e.g. isolated boxes or infinite lattices) and the dynamic processes (e.g. changing the sign of $J_4$ couplings). But we can always form two different frequency patterns  $\{\omega_{0,A}, \omega_{0,B}\}$ from the beginning and treat the potential deviations as small local detunings (as will be discussed in Sec.~\ref{sec:det}). 
Meanwhile, $\omega_{1,j}$ can be adjusted by parameters $L'_j$, $C'_j$ and $V_{AC,j}$ such that it is comparable to  $\omega_{0,j}$. 

\subsection{Generalized NMR protocol}
\label{sec:nmr}

In this section, we are going to present the core idea of our algorithm. The aim is to find a unitary gauge transformation $U(t)$ from ${\cal H}$ to $G$: $U(t) = \prod_{j} U_j(t) =  \prod_j e^{iF_j(t)}$, such that in the new gauge, the local magnetic field $\sigma_j^z$ vanishes and no additional couplings emerge. We denote $\psi(t)$ and $\phi(t)$ as  the eigenstates  of ${\cal H}$ and ${G}$ respectively. They are related by the transform $\phi(t) = U(t) \psi(t)$ and  $\phi(t)$ satisfy the  Schr{\" o}dinger equation $G \phi(t) = i\hbar \partial_t \phi(t)$. Therefore, $G = G_C + U{\cal H}_KU^{-1}, G_C =  \left( i\hbar \partial_t U \right) U^{-1} + U {\cal H}_C U^{-1}$.
Two of our requirements are as follows: (i) $G_C = 0$; (ii) $G = U{\cal H}_KU^{-1} = {\cal H}'_K$ where ${\cal H}_K'$ takes a similar Kitaev form with renormalized prefactors. We introduce the new variable $\tau_j = \omega_j t$ and we anticipate the test function $F_j = ({\alpha_j}/2)\left( \sin \tau_j \sigma_j^x - \cos \tau_j \sigma_j^y \right)$. By applying the mathematical steps in Appendix~\ref{sec:om}, from Eq.~(\ref{eq:gcexact}) we obtain
\begin{gather}
G_C = \frac{\hbar}{2} \sum_{j=1}^4 \left( \omega_{0,j} \cos \alpha_j + \omega_{1,j} \sin \alpha_j - \omega_j \cos \alpha_j + \omega_j \right)\sigma_j^z  \notag \\ 
- \left( \omega_{1,j} \cos \alpha_j - \omega_{0,j} \sin \alpha_j + \omega_j \sin \alpha_j\right) \left( \cos \tau_j \sigma_j^x + \sin \tau_j \sigma_j^y\right).
\label{eq:gco}
\end{gather}
The second time-dependent term vanishes for 
\begin{gather} 
\cos \alpha_j = - ({\omega_{0,j} - \omega_j})/{\sqrt{\omega_{1,j}^2 + \left( \omega_{0,j} - \omega_j\right)^2}}, \notag \\
\tan \alpha_j = {\omega_{1,j}}/{(\omega_{0,j} - \omega_j)}.
\end{gather}
$G_C$ then becomes a time-independent effective magnetic field polarized on $z$ direction only:
\begin{gather}
G_C = \sum_j \frac{\hbar}{2}\left( \omega_j - \sqrt{w_{1,j}^2 + \left( \omega_{0,j} - \omega_j \right)^2 } \right) \sigma_j^z.
\label{eq:gc0}
\end{gather}
If the frequencies of the AC voltages satisfy
\begin{gather}
\omega_j = \frac{ \omega_{1,j}^2 + \omega_{0,j}^2}{2\omega_{0,j}}, \qquad G_C = 0. \label{eq:wcondition}
\end{gather}

Next, we analyse the remaining part $U{\cal H}_KU^{-1}$ in the effective Hamiltonian $G$.
Constructed from spin operators, $U_j(t)$ commute between different sites. For the $\nu$-link ($\nu = x, y, z$), $U\sigma_{A}^{\nu} \sigma_{B}^{\nu}U^{-1} = ( U_{A}\sigma_{A}^{\nu} U_{A}^{-1}) (U_{B}\sigma_{B}^{\nu}U_{B}^{-1})$. In the rotating frame, from Eq.~(\ref{eq:si}) spin operators on each site undergo the following  gauge transformation:
\begin{gather}
\begin{split}
U_{j}\sigma_{j}^x U_{j}^{-1} &= \left( 1+ \cos^2 (\tau_j)  (\cos \alpha_j -1)\right)\sigma_j^x \\ &\phantom{=} + \frac{\cos \alpha_j -1}{2} \sin(2\tau_j )\sigma_j^y - \sin \alpha_j \cos(\tau_j)  \sigma_j^z,  \\
U_{j}\sigma_{j}^y U_{j}^{-1}& = \left( 1+ \sin^2(\tau_j)  (\cos \alpha_j -1)\right)\sigma_j^y \\ &\phantom{=} + \frac{\cos \alpha_j -1}{2} \sin(2\tau_j )\sigma_j^x - \sin \alpha_j \sin(\tau_j)  \sigma_j^z, \\
U_j \sigma_j^z U_j^{-1} &= \cos \alpha_j \sigma_j^z + \sin \alpha_j \cos (\tau_j) \sigma_j^x + \sin \alpha_j \sin (\tau_j) \sigma_j^y.
\end{split}
\label{eq:siu}
\end{gather}
{We denote $\left \langle f(t)\right \rangle_T $ as} the time average $({1}/{T}) \int_0^{T} f(t) dt$. Averaging over a long timescale $T = NT_A = T_B$ ($T_{j} = 2\pi/ \omega_j$, $N$ any integer larger than one), most of the time-dependent terms in the product  $( U_{A}\sigma_{A}^\nu U_{A}^{-1} ) ( U_{B}\sigma_{B}^\nu U_{B}^{-1} )$ will vanish. However, terms such as 
 $\left \langle  \cos^2(\tau_{A/B}) \right \rangle_T = \left \langle  \sin^2(\tau_{A/B}) \right \rangle_T = {1}/{2},
\left \langle \cos^2(\tau_A)  \cos^2(\tau_B)\right \rangle_T = \left \langle  \sin^2(\tau_A)  \sin^2(\tau_B)\right \rangle_T = {1}/{4}$ will remain. By imposing different frequency patterns for {sublattices} $A$ and $B$, we ensure that only Kitaev couplings are non-vanishing after the rotation
\begin{gather}
\left \langle G \right \rangle_T  = \left \langle U{\cal H}_K U^{-1}\right \rangle_T = {\cal H}'_K, \quad J'_{\nu} = r_{\nu} J_{\nu},
\label{eq:G}
\end{gather}
with $r_{\nu}$ ($\nu = x, y, z$) listed in Table~\ref{tab:nmr}.

\begin{table}[h]
\begin{center}
\caption{Parameters for generalized NMR protocol}  \label{tab:nmr}
\begin{tabular}{cc}
\hline
\hline
Parameter & Relation  \\
\hline
   $\alpha$ & $\arctan(2\omega_0\omega_1/(\omega_0^2 - \omega_1^2))$ \\
      $r_x, r_y$ & $\cos^2 \left( \alpha_A/2\right) \cos^2 \left( \alpha_B/2\right)$ \\ $r_z$ & $\cos \alpha_A \cos \alpha_B$ \\
   $u$ & $\cos \alpha_A - 1$ \\ $v$ & $\cos \alpha_B -1$ \\
   $r_1$ & $u^2v^2/64 + (u^2v + uv^2 + u^2 + v^2 )/8$ \\ & $ + uv + u + v + 1$\\ 
   $r_2$ & $u^2v^2/64$ \\
   $r_3$ & $u^2v^2/64 + (uv^2 + v^2)/8$ \\ $r_4$ & $u^2v^2/64 + (u^2v + u^2)/8$\\
     \hline
\hline
\end{tabular} 
\end{center}
\end{table}

\subsection{Measuring flux states through multi-channels}
\label{sec:pd}

Within a single box, we define four types of loop operators {in the rotating frame with Hamiltonian $G$ (\ref{eq:G})}:
\begin{gather}
\begin{split}
\mathcal{P}_c = \sigma_1^x\sigma_2^x\sigma_3^y\sigma_4^y &= c_1c_2c_3c_4, \\
\mathcal{P}_d = \sigma_1^y\sigma_2^y\sigma_3^x\sigma_4^x &= d_1d_2d_3d_4, \\
\mathcal{P}_e =\sigma_1^y \sigma_2^x \sigma_3^y \sigma_4^x &= -d_1c_2c_3d_4, \\
\mathcal{P}_f = \sigma_1^x\sigma_2^y\sigma_3^x\sigma_4^y &= -c_1d_2d_3c_4.
\end{split}
\end{gather}
These operators will be important in the detection of $\mathcal{Z}_2$ gauge fluxes. In particular, in the limit of strong horizontal bonds, as mentioned in the introduction we predict $\mathcal{P}_c=c_1 c_2 c_3 c_4=1$. 
In our Majorana representation (\ref{eq:maj}), they become four-body Majorana couplings. $\mathcal{P}_d = 1$ corresponds to the $\pi$-flux configuration while $\mathcal{P}_d = -1$ relates to the $0$ flux. The NMR protocol thus enables us to measure experimentally the flux states encoded in $\mathcal{Z}_2$ gauge fields. We denote $\left<  U \mathcal{P} U^{-1} \right>_T = \langle\langle {\cal P} \rangle\rangle$ as the time-averaged measurement (over the large Floquet period) in the original spin space. From Eq.~(\ref{eq:siu}), the unitary transformation to the rotating frame entangles these four loop operators
\begin{gather}
\begin{pmatrix}
\langle \langle \mathcal{P}_d \rangle \rangle \\
\langle \langle \mathcal{P}_c \rangle \rangle \\
\langle \langle \mathcal{P}_e \rangle \rangle \\
\langle \langle \mathcal{P}_f \rangle \rangle
\end{pmatrix} =
\begin{pmatrix}
r_1 & r_2 & r_3 & r_4 \\
r_2 & r_1 & r_4 & r_3 \\
r_3 & r_4 & r_1 & r_2 \\
r_4 & r_3 & r_2 & r_1
\end{pmatrix}
\begin{pmatrix}
\mathcal{P}_d \\
\mathcal{P}_c \\
\mathcal{P}_e \\
\mathcal{P}_f
\end{pmatrix}. \label{eq:mat}
\end{gather}
The coefficients read
\begin{gather}
\begin{split}
r_1 &= \left< \left( 1+ \sin^2(\tau_A)u\right)\cdot\left( 1+ \sin^2(\tau_B)v\right)\cdot \right. \\
&\phantom{=} \left. \left( 1+ \cos^2(\tau_B)v\right)\cdot\left( 1+ \cos^2(\tau_A)u\right)\right>_T, \\
r_2  &= \frac{u^2v^2}{16}  \left< \sin^2 (2\tau_A)\sin^2 (2\tau_B)\right>_T, \\
r_3 &= \frac{v^2}{4}\left<  \sin^2 (2\tau_B) \cdot \left( 1+ \sin^2(\tau_A)u \right) \cdot \left( 1+ \cos^2(\tau_A)u \right)\right>_T, \\
r_4  &= \frac{u^2}{4}\left<  \sin^2 (2\tau_A) \cdot \left( 1+ \sin^2(\tau_B)v \right) \cdot \left( 1+ \cos^2(\tau_B)v \right)\right>_T,
\end{split}
\end{gather}
where $u = \cos \alpha_A - 1, v = \cos \alpha_B - 1$. The time-averaged values of $r_i$'s are given in Table~\ref{tab:nmr}. Flux operators can be measured directly from the observables in the original frame by the inverse matrix in Eq.~(\ref{eq:mat}). For instance,
\begin{gather}
\mathcal{P}_d = \frac{1}{\mathcal{D}} \left( \tilde{r}_1 \langle \langle \mathcal{P}_d\rangle \rangle + \tilde{r}_2 \langle \langle \mathcal{P}_c\rangle \rangle + \tilde{r}_3 \langle \langle \mathcal{P}_e\rangle \rangle + \tilde{r}_4 \langle \langle \mathcal{P}_f \rangle \rangle \right), \label{eq:pd}
\end{gather}
where $\mathcal{D} = \sum_{m=1}^4 r_m^4 - 2\sum_{m<m'} r_m^2r_{m'}^2 + 8 \prod_{m=1}^4 r_m$ and 
$\tilde{r}_m = r_m \left( r_m^2 - \sum_{m' \ne m} r_{m'}^2\right) + 2 \prod_{m' \ne m}r_{m'}$. A similar formula is obtained for $\mathcal{P}_c$, through Eq. (27). 

\section{Numerical test}

\subsection{Time-averaged quantities}

We test the protocol (valid to any order in $1/\omega_j$) numerically by solving the time-dependent Hamiltonian with a diagonalization using Julia scientific computing language and we evaluate the time-averaged observables $\langle\langle \sigma_j^z\rangle\rangle$ and $\langle\langle \sigma_j^z \sigma_l^z\rangle\rangle$. We choose different integer values $N=3,5,7$ and check that the results are (almost) identical. Here, $\langle\langle f\rangle\rangle=\langle\langle f\rangle(t)\rangle_T$ denotes the time averaged quantity $(1/T)\int_0^T \hbox{Tr}(\rho(t) f)$ with $\rho(t)$ being the density matrix of the system and $T= 2\pi/\omega_{min}$ with $(\omega_{min}= {\omega_B})$. Therefore, $T$ {corresponds} to the largest Floquet period. 

\begin{figure}[t]
  \begin{center}
     \includegraphics[width=0.4\textwidth]{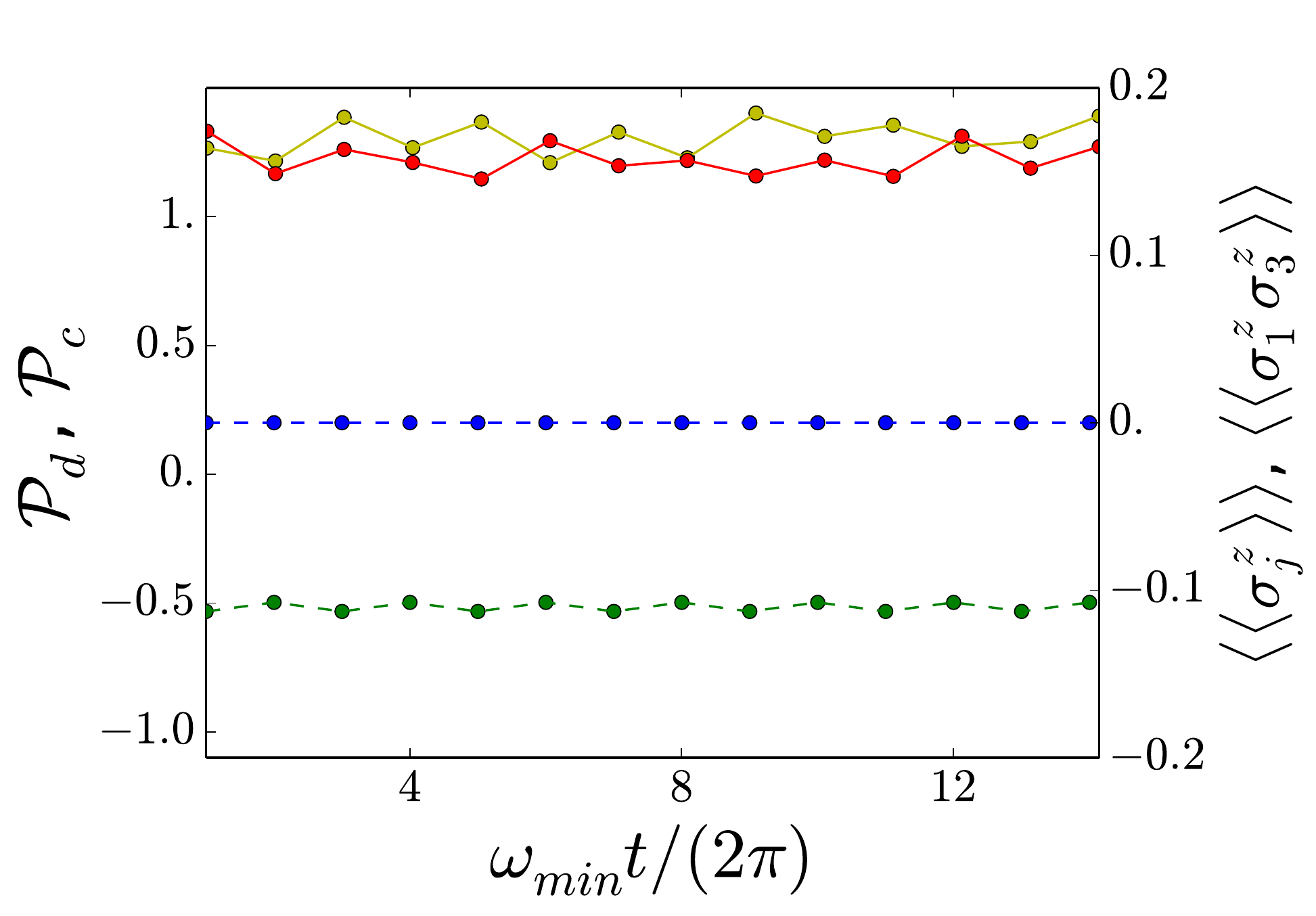}\\
  \end{center}
  \vskip -0.5cm \protect\caption[]
  {(color online) Time evolution of $\langle\langle \sigma_j^z\rangle\rangle$ (blue) and $\langle \langle \sigma_1^z \sigma_3^z \rangle\rangle$ (green) (dashed lines); and of the fluxes $\mathcal{P}_d$ (yellow) and $ \mathcal{P}_c $ (red) ({solid} lines) averaged over the longest period $2\pi/\omega_{min}$ with $\omega_{min}={\omega_A/N = \omega_B}$. We took $N=3$, but other integer values of $N$ give comparable results. The NMR frequency pattern is selected on each site as $\omega_{1,j} = \sqrt{2} \omega_{0,j}$, $\omega_j = 3\omega_{0,j}/2$. {(These initial frequency conditions remain the same in Figs.~\ref{fig:nmr} - \ref{fig:nmr_impurity})} The top panel corresponds to weak vertical bonds $|J_1|=|J_2|=0.4 {\hbar \omega_B}$, $|J_3|=|J_4|=0.045 |J_1|$, while the bottom {panel} deals with the regime of strong vertical bonds $J_3/\epsilon_3=J_4/\epsilon_4=0.8$.}
  \label{fig:nmr0}
  \vskip -0.5cm
\end{figure}

The calculation of spin observables averaged in time {under the Hamiltonian $\mathcal{H}$} should agree with the calculation in the rotating frame with the Hamiltonian $G$. In Fig.~\ref{fig:nmr0}, we show results in the particular limit of strong vertical bonds with antiferromagnetic couplings $J_3=J_4\gg |J_1|=|J_2|$. We verify $\langle\langle \sigma_j^z\rangle\rangle=0$ since on each site a spin can be polarized in the $|+z\rangle$ and $|-z\rangle$ direction equally. We check that $\langle\langle \sigma_j^x\rangle\rangle$ and $\langle\langle \sigma_j^y\rangle\rangle$ are zero. In Fig.~\ref{fig:nmr0}, we check the correct value $\langle\langle \sigma_1^z\sigma_3^z\rangle\rangle \sim -1\times r_z = -0.11$ (due to the large $J_3$ coupling in the rotating frame). 

We can also detect directly the flux variables through the 4-body spin operators and compare with the mathematical predictions above. In Fig.~\ref{fig:nmr0}, we show that we obtain numerically in the regime of weak vertical bonds $\mathcal{P}_c\sim \mathcal{P}_d \sim  1$ {from the measurement of four separate channels $\langle \langle \mathcal{P}_\xi \rangle \rangle$ ($\xi = c, d, e, f$)}, using formulas (26) and (27),  corresponding to the precise engineering of the $\pi$-flux configuration.

\subsection{Detuning effects}
\label{sec:det}

We have three steps of fine tunings throughout our proposal:
(i) The cancellation of vertical $X$ couplings;
(ii) The engineering of a circularly polarized NMR field {in Hamiltonian (\ref{eq:hf})};
(iii) The cancellation of local magnetic field in the rotating frame.
The prerequisite (i) is important for the realization of Kitaev type Hamiltonians. We show in Sec. IIID that such perturbations can be useful to produce local flux impurities, at a perturbation level.

For (i), the condition for the parameters from Eq.~(\ref{eq:xx}) becomes
\begin{gather}
E_{\widetilde{L},m} = -E_{J,m}/2, \quad m = 3, 4.
\end{gather}
This can be reached by tuning the phases $\Phi_3, \Phi_4$. We {will} discuss this point more carefully in Sec.~\ref{sec:per}.

For (ii), we impose $\omega_{1,j} = \omega_{L',j} = \omega_{C',j}$ in terms of parameters {(see Table III in Appendix~\ref{app:para})}. 
We discuss below perturbation effects from that condition.

Now for the algorithm (iii), we consider a small deviation in the frequency pattern $\omega_j \rightarrow \widetilde{\omega}_j  = \omega_j + \delta \omega_j$. The Hamiltonian of the NMR field becomes
\begin{gather}
 {\cal H}_{\text{NMR}}(t) = - \sum_{j=1}^4 \frac{\hbar\omega_{1,j}}{2}(\cos (\widetilde{\omega}_j t) \sigma_j^x + \sin (\widetilde{\omega}_j t)\sigma_j^y)\notag \\
 + \frac{\hbar\omega_{1,j}}{2} \frac{\delta \omega_j }{\omega_j} \cos (\widetilde{\omega}_j t) \sigma_j^x. \label{eq:hacdt}
 \end{gather}
 The third term is also equivalent to change $ \omega_{L',j}$ while $\omega_{C',j}$ remains unchanged in relation with Eq. (19). More details on the parameters of the box are given in Appendix A.  
We can study the consequences of the detuned Hamiltonian $(\ref{eq:hacdt})$ in the {rotating} frame. Firstly,  the variable $\widetilde{\alpha}_j$ characterizing the unitary transformation has a small shift:
 \begin{gather}
 \begin{split}
 \cos \widetilde{\alpha}_j & \simeq \cos \alpha_j + \frac{\cos \alpha_j (1-\cos^2 \alpha_j)}{1- \omega_{0,j}/\omega_j} \frac{\delta \omega_j}{\omega_{j}}, \\
 \sin \widetilde{\alpha}_j & \simeq \sin \alpha_j -\frac{\cos^2 \alpha_j}{1- \omega_{0,j}/\omega_j}  \frac{\delta \omega_j}{\omega_{j}}.
 \end{split}
 \end{gather}
 When $\delta \omega_j \ll \omega_j$, we can assume $ \cos \widetilde{\alpha}_j \simeq \cos \alpha_j, \sin \widetilde{\alpha}_j \simeq \sin \alpha_j$. The effective Hamiltonian $G_C$ {in Eq.~(\ref{eq:gc0})} takes the form accordingly
 \begin{gather}
G_C  \simeq \sum_j \frac{\hbar\omega_{0,j}}{2\omega_{j}} \delta \omega_j \sigma_j^z.
\label{eq:con1}
\end{gather}
In our numerical simulation $\omega_0 \sim \omega$, $G_C$ becomes sensitive under detuning. To analyze the {consequence} of the extra {third} term in the Hamiltonian $(\ref{eq:hacdt})$, we go back to the general unitary transform (\ref{eq:siu}) and after time average
\begin{gather}
\left< \left< \frac{\hbar\omega_{1,j}}{2} \frac{\delta \omega_j }{\omega_j} \cos (\widetilde{\omega}_j t) \sigma_j^x \right>\right> \simeq  \frac{\hbar}{4} \left( \frac{2\omega_{0,j}}{\omega_j} - \frac{\omega^2_{0,j}}{\omega^2_j} \right)\delta \omega_j \sigma_j^z,
\label{eq:con2}
\end{gather}
where we keep the initial large time period $T(\omega)$ unchanged and $\left< \cos^2(\widetilde{\omega}_jt) \right>_T \simeq 1/2 + \mathcal{O}(\delta \omega_j)$. In the end, {combining Eqs.~(\ref{eq:con1}) and (\ref{eq:con2})} we expect the detuning $\omega_j + \delta \omega_j$ on each site would create a non-zero effective magnetic field: 
\begin{gather} 
{\cal \widetilde{H}}_z  = \sum_{j} \frac{\omega_{0,j}}{\omega_j} \left( 1-\frac{\omega_{0,j}}{4\omega_j}\right) \hbar \delta \omega_j \sigma_j^z. \label{eq:hzdt}
\end{gather}
The pre-factor cannot be zero, otherwise $\omega^2_{1,j} < 0$ by the  relation (\ref{eq:wcondition}): $2\omega_j\omega_{0,j} = \omega^2_{1,j} + \omega^2_{0,j}$. The gapped phase is protected to the first order perturbation under $\widetilde{H}_z$. To second order $\mathcal{O}(\delta\omega/|J_1|)$, effective couplings $\sigma_1^z\sigma_2^z$ and $\sigma_3^z\sigma_4^z$ are generated but quite small. For the gapless phase (e.g. in the Kitaev honeycomb model), the magnetic field is polarized purely along $z$ direction without a gap opening.

\begin{figure}[t]
  \begin{center}
     \includegraphics[width=0.4\textwidth]{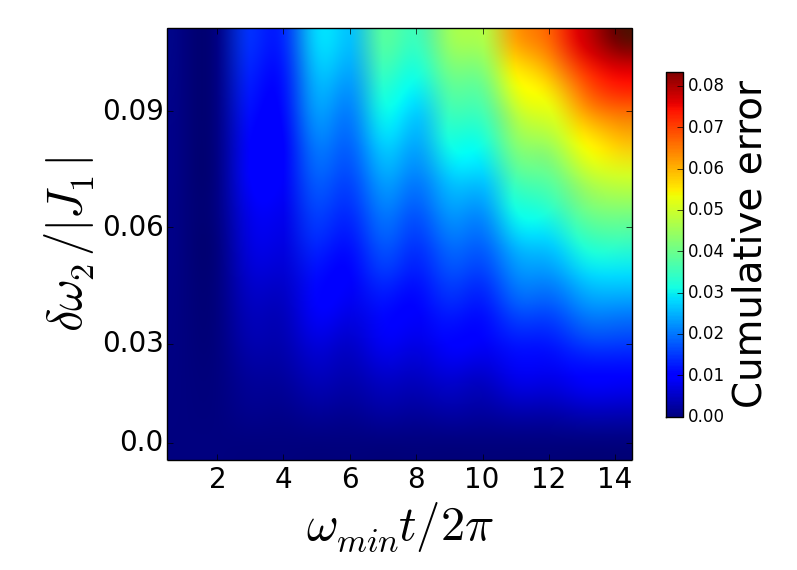}
  \end{center}
  \vskip -0.5cm \protect\caption[]
  {(color online) Detuning effects in $\delta\omega_2$ of the driving frequency $\omega_2$. Average error on {$\mathcal{P}_d$} (averaged over time) induced by this detuning, as a function of both $\delta \omega_2$ and the adimensional time $\omega_{min}t/2\pi$. The errors are relatively small, one gets errors of less than $3\%$ for more than 14 time periods, if the detuning is of the order of $5\%$.  This plot corresponds to the weak vertical bonds configuration (see Fig.   \ref{fig:nmr0}).
}
  \label{fig:nmr}
  \vskip -0.5cm
\end{figure}

Numerically, we check the above effects by simultaneously detuning four sites or a single site.  As a numerical test, we show results on detuning  $\delta \omega_j$ compared to  $\omega_j$. 
All physical observables  (especially $\mathcal{P}_d$) are supposed to be stable via a small detuning. When $\delta \omega_j$ is comparable to  $\omega_j$, we could detect large fluctuations. In Fig.~\ref{fig:nmr}, we show the effect of detuning the driving frequency of the site $2$ on the gauge-field four-body operator ${\cal P}_d$. We check that one gets small errors of the order of $3\%$ for more than 14 time periods if the detuning is of the order of $5\%$. 

\subsection{Dissipative processes}

\begin{figure}[t]
  \begin{center}
     \includegraphics[width=0.4\textwidth]{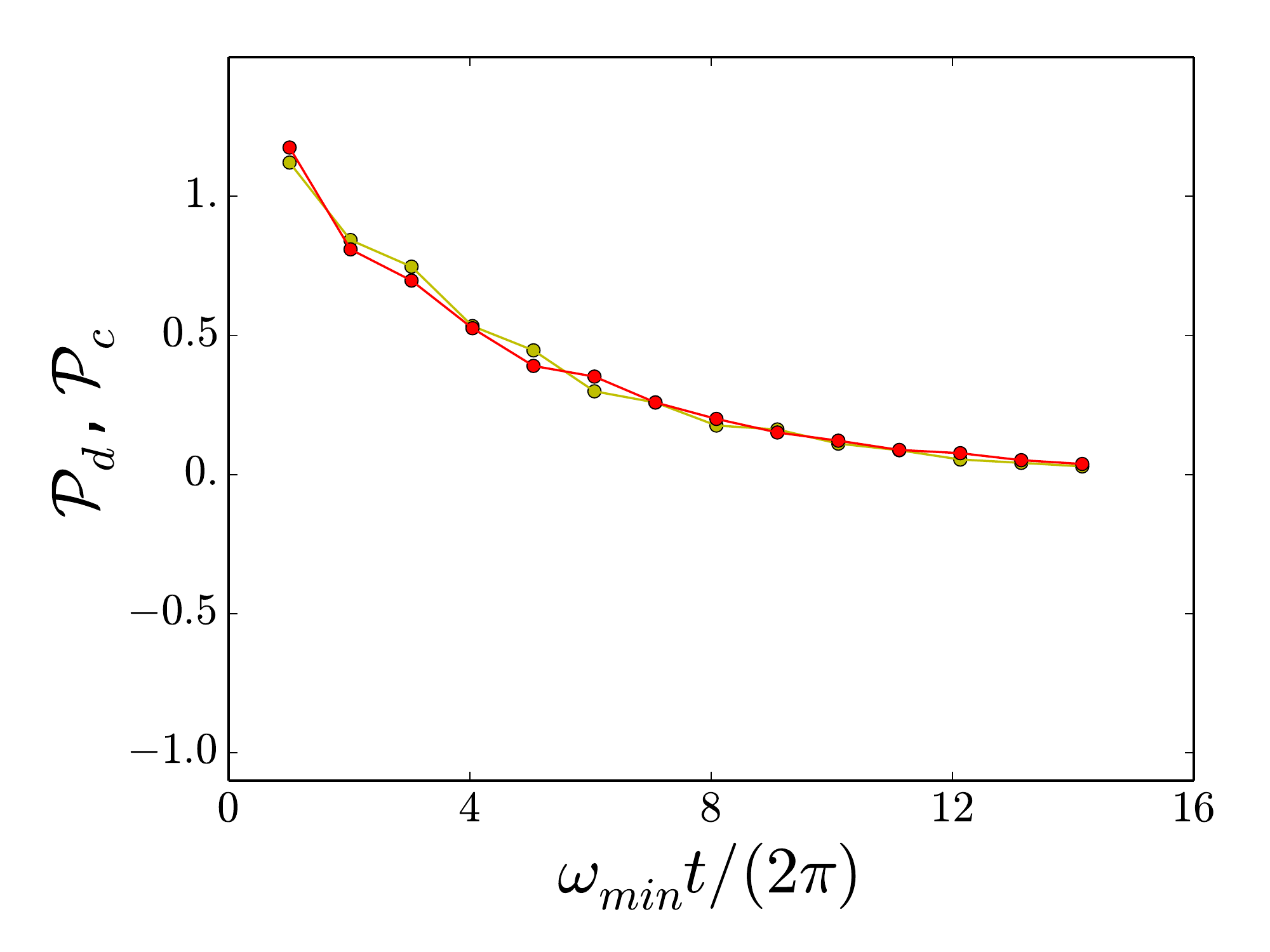}
  \end{center}
  \vskip -0.5cm \protect\caption[]
  {(color online) Time evolution of the fluxes $\mathcal{P}_d$ (yellow) and $ \mathcal{P}_c $ (red) {in dissipative processes}. Here, we have taken weak vertical bonds $|J_1|=|J_2|=0.4 {\hbar \omega_B}$, $|J_3|=|J_4|=0.045 |J_1|$. Losses and dephasing, with rates $\Gamma= 5~10^{-3} {\omega_B}$ and $\gamma= 5~10^{-3} {\omega_B}$, lead to a monotonous exponential decay of the fluxes  $\mathcal{P}_d$  and $ \mathcal{P}_c $ from their initial quantized value $+1$ to zero. Results of this figure must be compared with those of Fig. 3.}
  \label{fig:fluxdissipative}
  \vskip -0.5cm
\end{figure}

It is important to characterize the influence of losses and dephasing on the dynamical protocols.
Taking into account these physical processes, the dynamics of the qubit density matrix $\rho$ is described by the following Lindblad-type master equation,
\begin{align}
\partial_t \rho = &-(i/\hbar) \left[{\cal H}(t), \rho \right] + \gamma \sum_{j=1}^4 \left(\sigma_j^z \rho \sigma_j^z - \rho\right) \notag \\
&+\frac{\Gamma}{2} \sum_{j=1}^4 \left( 2 \sigma_j^+ \rho \sigma_j^- - \sigma_j^+ \sigma_j^- \rho - \rho \sigma_j^+ \sigma_j^-\right).
\end{align}
Here ${\cal H}(t)$ is the original time-dependent Hamiltonian in Eq.~(\ref{eq:hf}) and $\gamma$ and $\Gamma$ are respectively the dephasing and loss rates of the qubit; we suppose independent losses and dephasing on each site, with the same strength. As can be seen in Fig. \,\ref{fig:fluxdissipative}, the presence of losses and dephasing {destroys} the quantization of both $\mathcal{P}_d$ (yellow) and $ \mathcal{P}_c $ (red) at the level of one box. Studying the effect of dephasing and losses separately, we find that they lead qualitatively to a similar decay in the flux dynamics. When simulating the proper Hamiltonian in an experiment, one should therefore perform all measurements within a timescale $\tau_{mes}$ set by these characteristic rates, $\tau_{mes}\ll 1/\gamma,1/\Gamma$. It is relevant to note the similar role $\gamma$ and $\Gamma$ in these measurements. 

\subsection{Perturbations and Changing {fluxes}}
\label{sec:per}

Here, we analyze the effects of non-zero vertical $X$ couplings on single-box systems, arising from Josephson junctions. In the limit of strong horizontal bonds, the ground state is highly degenerate: $\left| \text{GS} \right> = \left| \alpha \alpha \right>_{x, (1,2)} \otimes  \left| \beta \beta \right>_{y, (3,4)}, (\alpha, \beta) = \pm 1$. From perturbation theory, interactions on the vertical bonds {contribute to} ${\cal H}_{\text{eff}}^{(2)}   = - {J_3J_4}/{(|J_1| + |J_2|)} \left( \sigma_1^z\sigma_2^z \sigma_3^z\sigma_4^z \right)_{\text{eff}} - {J_3^xJ_4^x}/{|J_2|} \left( \sigma_1^x\sigma_2^x \sigma_3^x\sigma_4^x \right)_{\text{eff}}$.
Strong $J_1$ links ensure that $\left< \sigma_1^x\sigma_2^x \right> = 1$. Thus, 
  \begin{gather}
{\cal H}^{(2)} = - \frac{J_3J_4}{|J_1| + |J_2|} \left< \sigma_1^z\sigma_2^z \sigma_3^z\sigma_4^z \right>  - \frac{J_3^xJ_4^x}{|J_2|} \left< \sigma_3^x\sigma_4^x \right>. \label{eq:j3x}
  \end{gather}
  In the Majorana basis {(\ref{eq:maj})},
  \begin{gather}
  \left< \sigma_1^z\sigma_2^z \sigma_3^z\sigma_4^z \right> = \mathcal{P}_c \mathcal{P}_d = \mathcal{P}_d, \quad \left< \sigma_3^x\sigma_4^x \right> = -id_3d_4,
  \label{eq:j3x_1}
      \end{gather}
      where we have taken into account $\mathcal{P}_c =  \left< \sigma_1^x \sigma_2^x \sigma_3^y \sigma_4^y \right> = 1$. 
      
 Once we add an additional inductance $\widetilde{L}_3$ between sites $1$ and $3$ and turn off the vertical $X$ coupling such that $J_3^x + \widetilde{J}_3^x = 0$ (we have $\Phi_3$ fixed and $J_3 > 0$), the contribution from $J_4^x$ vanishes and we check that $\sigma_2^x\sigma_4^x$ becomes an irrelevant operator to any higher order in perturbation theory. The gapped phases of Kitaev type spin models are therefore fully protected against local $J_4^x$ noises. {This point is crucial to the flux engineering later in Sec.~\ref{sec:flux1}}. 
             
Furthermore, we gain the flexibility of tuning the $\Phi_4$ phase, which is useful to engineer local defects with 0 flux in a unit cell.  Suppose we deviate from the condition in {Eq.~(\ref{eq:xx})}, and study some effects of $J_3^x$ and $J_4^x$.  To second-order in $J_3^x J_4^x$, we then engineer a term in the Hamiltonian, which is equivalent to add a small inductance between the sites $3$ and $4$: $\delta {\cal H}_{\parallel} = \delta J_1 \sigma_3^x \sigma_4^x=- i \delta J_1 d_3 d_4$, where $\delta J_1$ is proportional to $J_3^x J_4^x$.  Tuning progressively the flux $\Phi_4$ in time {would change the sign of $J_4^x$ from positive to negative}. Then this allows us to locally change the flux in a square cell {from $\pi$ to $0$} and have also a time control on the local gauge fields.  {Next} we discuss this protocol in more detail. 

\begin{figure}[t]
  \begin{center}
     \includegraphics[width=0.4\textwidth]{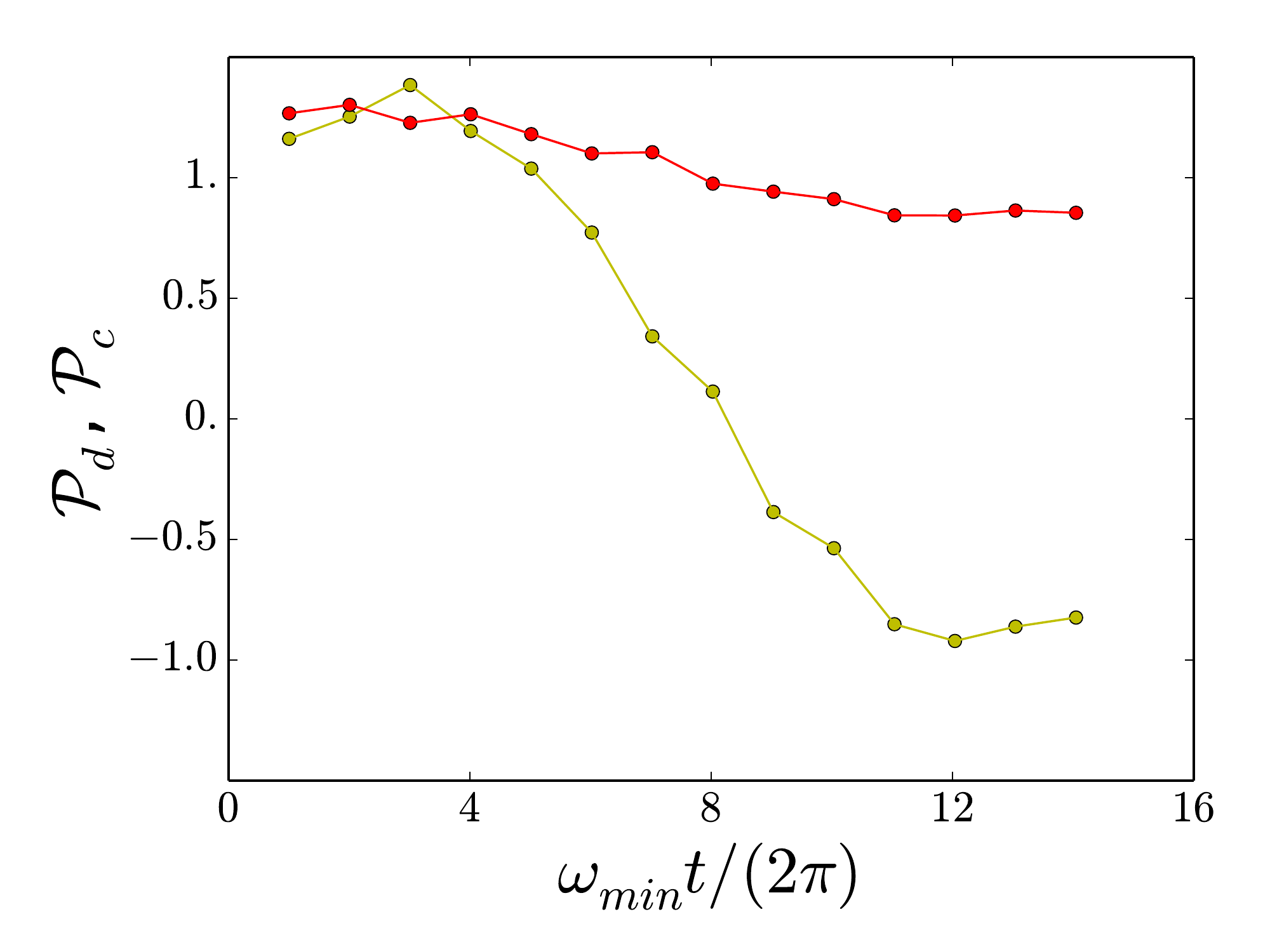}
  \end{center}
  \vskip -0.5cm \protect\caption[]
  {(color online) Time evolution of $\mathcal{P}_d $ (yellow) and $\mathcal{P}_c$ (red) {under a parity flip}. Here, we have taken weak vertical bonds $|J_1|=|J_2|=0.4 {\hbar \omega_B}$, and $2 |J_3|=|J^x_3|=0.1 |J_1|$. We have considered a sinusoidal variation of $2J_4=J^x_4$ between {the range} $\pm 0.1 |J_1|$. {An additional small field $h_y \sigma_3^y$ is implemented} with $h_y=0.08 ÍJ_1Í$.}
  \label{fig:nmr_impurity}
  \vskip -0.5cm
\end{figure}

In this protocol,  we flip the sign of the parity operator $-i d_3 d_4$ in time.  The ground state of ${\cal H}_K + \delta {\cal H}_{\parallel}$ differs depending on the sign of $\delta J_1$ ({or $J_3^xJ_4^x$} which could be tuned by some local magnetic flux {like $\Phi_4$}), corresponding to the two choices of the parity operator $i d_3 d_4$ ($+1$ or $-1$). In order to make such a protocol, one needs to avoid a gap closing when $\delta J_1=0$ because the system would not follow adiabatically the required ground state. Therefore, this {dynamical} protocol also requires an additional small field $h_y \sigma_3^y$  coupling the two ground states. Such a term physically can be derived by analogy {with} the NMR device, by coupling locally the site $3$ capacitively to a small DC constant bias voltage.  One can {then} control the strength of $h_y$ in this case since it is proportional to the capacitance and to the bias voltage. This precise time-control on local fluxes is illustrated in Fig.~\ref{fig:nmr_impurity}, where $\mathcal{P}_d $ is progressively changed from +1 to -1 while $\mathcal{P}_c$ remains roughly constant. We already observe this effect without using optimized geodesic paths \cite{Tomka}. 
 
\section{Application In coupled-box ensembles}

\subsection{Quantum spin liquids, Majorana states, Probes}

In the two-dimensional lattice of Fig.~\ref{fig:lattice},  once a box unit cell is built up one can construct more complex geometries with $J_4 \neq 0$ for square ladders \cite{KFA}, $J_4  = 0$ for brick-wall ladders \cite{KFA} and their equivalents in two dimensions, the Kitaev honeycomb model \cite{Kitaev}.  The three gapped  spin-liquid phases $A_x$, $A_y$, $A_z$ (with short-range entanglement emerging in the $X$, $Y$ and $Z$ directions) and the gapless $B$ phase in these spin models could be observed. In the Kitaev honeycomb lattice, the $A_z$ gapped phase supports a toric code \cite{Kitaevcode} and the $B$ phase allows non-Abelian anyonic statistics in the presence of a magnetic field.  It is important to mention recent efforts in quantum materials to observe through Nuclear Magnetic Resonance the gap in the $B$ phase opening in the presence of magnetic fields as well as topological aspects through neutral edge mode measurements \cite{Klanjsek,thermaltransport}. 
One could also envision to build `decorated' ladders showing chiral spin liquid states \cite{Yao}.

In addition, the Kitaev spin chain can be mapped to the transverse field Ising model and the two-leg square ladders have the dual of the XY chain in alternating transverse fields \cite{KFA, Feng}. 
Spin-spin correlation functions could reveal the short-ranged entanglement in gapped phases \cite{Roushan}. Here, we discuss how the NMR device can be used to detect Majorana physics and quantum phase transitions in Kitaev spin models. 

Let us assume the quantum phase transition with decoupled (zig-zag) chains in the two-dimensional honeycomb lattice model, $J_3=J_4=0$. In Fig.~\ref{fig:3box} (a), the quantum phase transition occurs when $\delta J_2=J_1$ {for the upper chain}. At the quantum phase transition, the Hamiltonian can be written in terms of Dirac fermions in the continuous limit by recombining {$c_{2m-1}$ and $c_{2m}$ along the chain}. The continuum model is a one-dimensional fermion Dirac model of 
$\psi(x)$ and  $\psi^{\dagger}(x)$ operators \cite{KFA} and spin-spin correlation functions show power-law decay. To probe the quantum critical fluctuations in the chain, one can weakly couple this chain to a spin S=1/2 $\vec{S}$ described by a transmon qubit, or another spinless fermion, that also reveals two Majorana fermions $c$ and $d$, such that $S_z=idc$, $S_x=c$ and $S_y=d$.  Adding a small coupling between this chain and the impurity spin (either capacitive or inductive {depending} on the location of this impurity spin), then one can engineer a small coupling $i\alpha d c_i$, where $\alpha\ll J_1$, involving the Majorana fermion $c_i$ at site $i$. By analogy to the two-channel Kondo model at the Emery-Kivelson line \cite{EK}, we identify a coupling term {$\propto$} $i\alpha d(\psi(x)+\psi^{\dagger}(x))$.  

The fermion $d$ will entangle with the chain and the Majorana fermion $c$ will remain free. A signature of this free remnant Majorana fermion is a $(\ln 2)/2$ entropy as well as a logarithmic  magnetic susceptibility $\chi_{imp}=\partial \langle S_z\rangle/\partial h\propto \ln h$, in contrast to a linear behavior for the one-channel Kondo model \cite{EK}.  With the NMR device attached to the spin-1/2 impurity, one could control the field {strength $hS_z$ by detuning the on-site frequency $\omega$ from Eq.~(\ref{eq:hzdt})} and measure the logarithmic growth of the susceptibility reflecting the Majorana physics as well as quantum critical fluctuations in the chain.
The gapped phases of the Kitaev model in ladder geometries also reveal edge mode excitations \cite{KFA}.  The NMR device could also probe in that case the susceptibility at low fields to detect these modes (A precise time-dependent protocol including perturbation effects for such a chain device will be studied in a further publication). These results do not probe non-Abelian statistics
\cite{Ivanov,MooreRead}, but still would give some response  of Majorana fermions. 

Boxes in the limit of strong vertical bonds could give rise to spin-1 quantum impurity physics \cite{KarynPhD}. 

\subsection{$\mathcal{Z}_2$ gauge fields and N\' eel order of {fluxes}}
\label{sec:flux1}
\begin{figure}[t]
  \begin{center}
    \includegraphics[width=8.6cm]{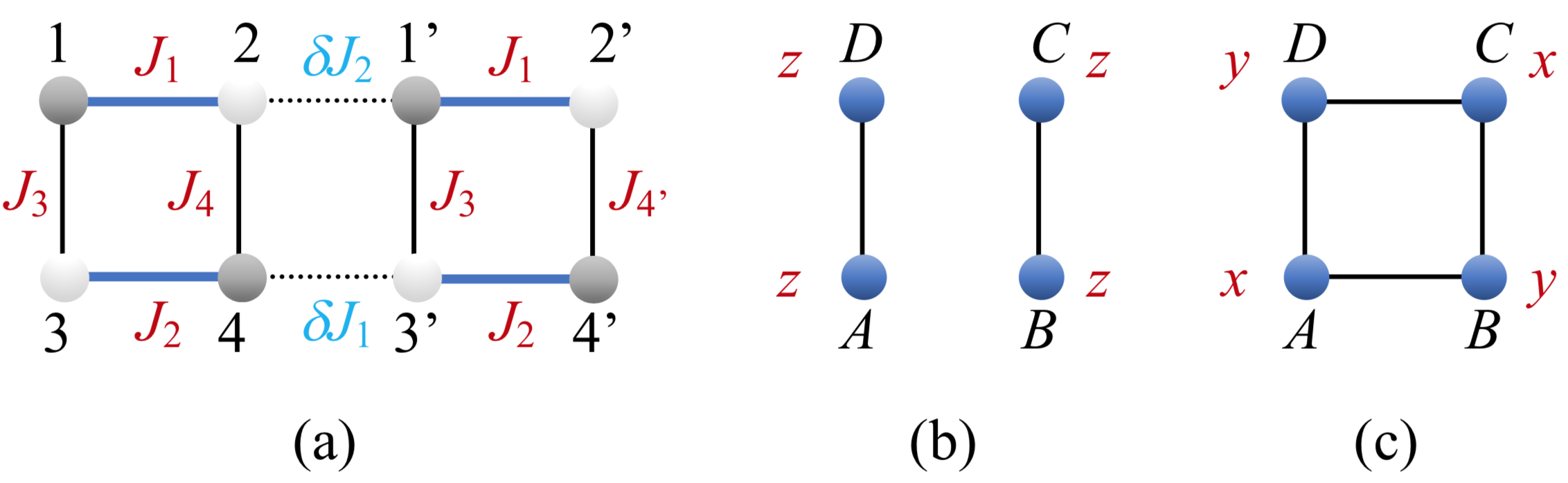}
          \end{center}
  \vskip -0.5cm \protect\caption[]
  {(color online) (a) Two coupled boxes in the limit of large $J_1$ and $J_2$; (b)-(c) Space of four effective spins formed by strong  $J_1$ and $J_2$ links; While non-zero $J_4$ and $J_{4'}$ reproduce Ising couplings (b), suppressing $J_4$ and $J_{4'}$ would lead to a four-body Hamiltonian (c) related to Wen's toric code. }
    \label{fig:3box}
    \vskip -0.5cm
\end{figure}

Now we discuss a peculiar limit of coupled-box systems, where inside each box all $c$ Majorana fermions  are gapped due to the large $J_1$ and $J_2$ couplings (shown in Fig.~\ref{fig:lattice} Right bottom). 
By coupling two boxes in the way of Fig.~\ref{fig:3box} (a) {with $J_3^x = 0$ and  $J_3 > 0$}, we are able to realize a N\'{e}el state of $d$-Majorana gauge fields. Performing perturbation theory in the spin space (see Appendix~\ref{sec:pert}) and mapping into the Majorana representation, we find:
\begin{gather}
\begin{split}
{\cal H}_{\text{eff}}^{(2)}  = \text{cst} &- \frac{J_3}{|J_1| + |J_2|}  \left( J_4\mathcal{P}_1 + J_{4'} \mathcal{P}_3 \right),  \\
{\cal H}_{\text{eff}}^{(4)} = \text{cst} &- \frac{J_3}{2(|J_1|+|J_2|)^3} \left( 2J_3J_4J_{4'} \mathcal{P}_{1} \mathcal{P}_{3}\right. \\
&\left.+ J_4(J_3^2+ J_{4'}^2) \mathcal{P}_1 + J_{4'}(J_3^2+ J_{4}^2)\mathcal{P}_3\right)  \\
&- \frac{\delta J_1 \delta J_2 }{2(|J_1|+|J_2|)^3} \left( 5J_3 J_{4'} \mathcal{P}_{\widetilde{123}} \right.\\
&\left.+J_3J_4 \mathcal{P}_2 + J_3^2 \mathcal{P}_{\widetilde{12}}+ J_4 J_{4'} \mathcal{P}_{\widetilde{23}}\right), \label{eq:ho}
\end{split}
 \end{gather}
 where $\mathcal{P}_{\mu}$ describes the four-body $d$-Majorana coupling on the vertices of box $\mu=1,2,3$ (in Fig.~\ref{fig:3box} (a), $\mu = 2$ denotes an induced box in the middle). More precisely,
 $\mathcal{P}_1 = d_1d_2d_3d_4, \quad \mathcal{P}_2 = d_2d_{1'}d_{4}d_{3'}, \quad \mathcal{P}_3 = d_{1'}d_{2'}d_{3'}d_{4'},
  \mathcal{P}_{\widetilde{12}} = \mathcal{P}_1 \mathcal{P}_2 = d_1d_{1'}d_3d_{3'}, \quad  \mathcal{P}_{\widetilde{23}} = \mathcal{P}_2 \mathcal{P}_3 = d_2d_{2'}d_4d_{4'},
  \mathcal{P}_{\widetilde{123}} = \mathcal{P}_1 \mathcal{P}_2 \mathcal{P}_3 = d_1d_{2'}d_3d_{4'}$. To minimize the energy, fluxes within each box can be uniquely fixed by the signs of $J_4$ and $J_{4'}$. 
 From the discussion of Sec.~\ref{sec:per}, we infer that when $J_3^x = 0$, non-zero $J_4^x$ and $J_{4'}^x$ couplings are allowed and do not enter into effective terms in any order of perturbation. Thus, the flexibility on the signs of $J_4$ and $J_{4'}^x$ is virtually guaranteed. In Table~\ref{tab:cbf}, we list all possible orderings of three gauge fields for two coupled boxes. 
  
  In large networks, one could couple more boxes in the same way and build square ladders. When all products of $J_3J_4$ are kept positive, the emergent $\pi$-flux ground state leading to the N\' eel order of 
 ${\mathcal Z}_2$ gauge fields is in agreement with Lieb's theorem. The N\' eel order could reveal a finite critical temperature in the case of long-range coupling between boxes, by analogy {with} the Ising model (see Sec.~\ref{sec:syk} below). By tuning the signs of $J_{4}$ one is able to create impurities of $0$ fluxes in the static $\mathcal{Z}_2$ gauge fields: a pair of fluxes in the bulk or a single flux on the boundary.  Another proposal to engineer many-body phases of fluxes in ladder systems has been done recently \cite{Alex}. {Small ladder spin systems generally reveal rich dynamics} due to Mott physics and gauge fields \cite{Review2018}.  From Eqs.~(\ref{eq:j3x})-(\ref{eq:j3x_1}), a small non-zero $J_3^x$ on the vertical $J_3$-links would fix the parity of two Majorana pairs $-id_3d_4$ and $-id_{3'}d_{4'}$, and would then help in deciding between the two possible ordered ground states with {$0$ or} $\pi$ order. 

 \begin{table}
\begin{center}
\caption{Ordering of gauge fields for two coupled boxes} 
 \label{tab:cbf}
  \begin{tabular}{ccc}
    \hline
\hline
 $\left(\text{sgn}[J_4], \text{sgn}[J_{4'}] \right)$& \phantom{flux} $(\mathcal{P}_1, \mathcal{P}_2, \mathcal{P}_3,  \mathcal{P}_{\widetilde{12}},  \mathcal{P}_{\widetilde{23}},  \mathcal{P}_{\widetilde{123}}) \phantom{flux} $  &  flux \\
 \hline
$(+,+)$ & $(+1, +1, +1, +1, +1, +1)$ & $\pi \quad \pi \quad  \pi$ \\
$(-,-)$ & $(-1, -1, -1, +1, +1, +1)$ & $0 \quad 0 \quad  0$ \\
$(+,-)$ & $(+1, +1, -1, +1, -1, -1)$ & $\pi \quad \pi \quad  0$ \\
$(-,+)$ & $(-1, -1, +1, +1, -1, -1)$ & $0 \quad 0 \quad  \pi$ \\
  \hline
\hline
\end{tabular}
\end{center}
\end{table}

\subsection{Towards Wen's toric code}
\label{sec:wen}

\begin{figure}[t]
  \begin{center}
    \includegraphics[height=2.9cm]{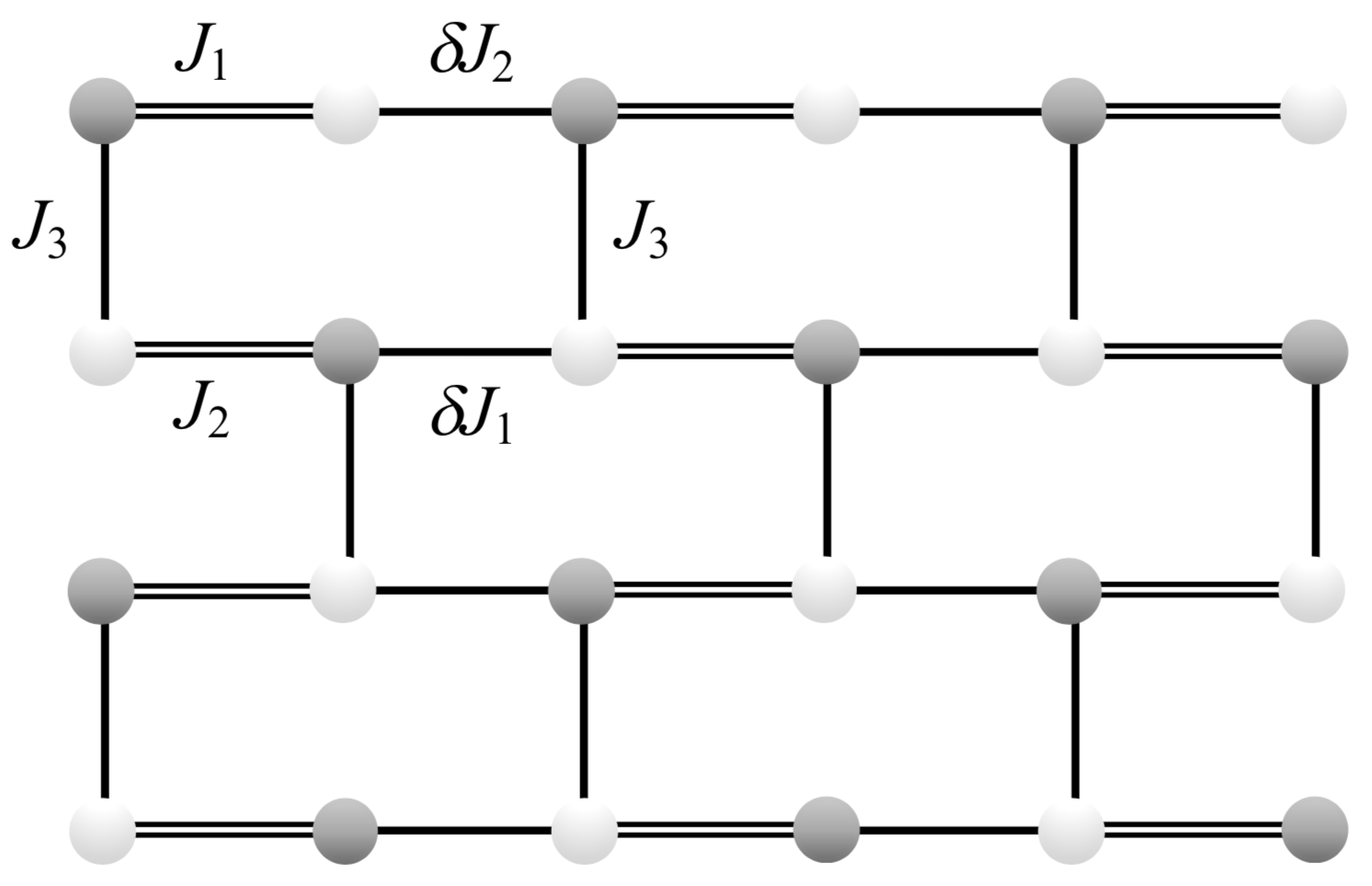} \    \includegraphics[height=2.9cm]{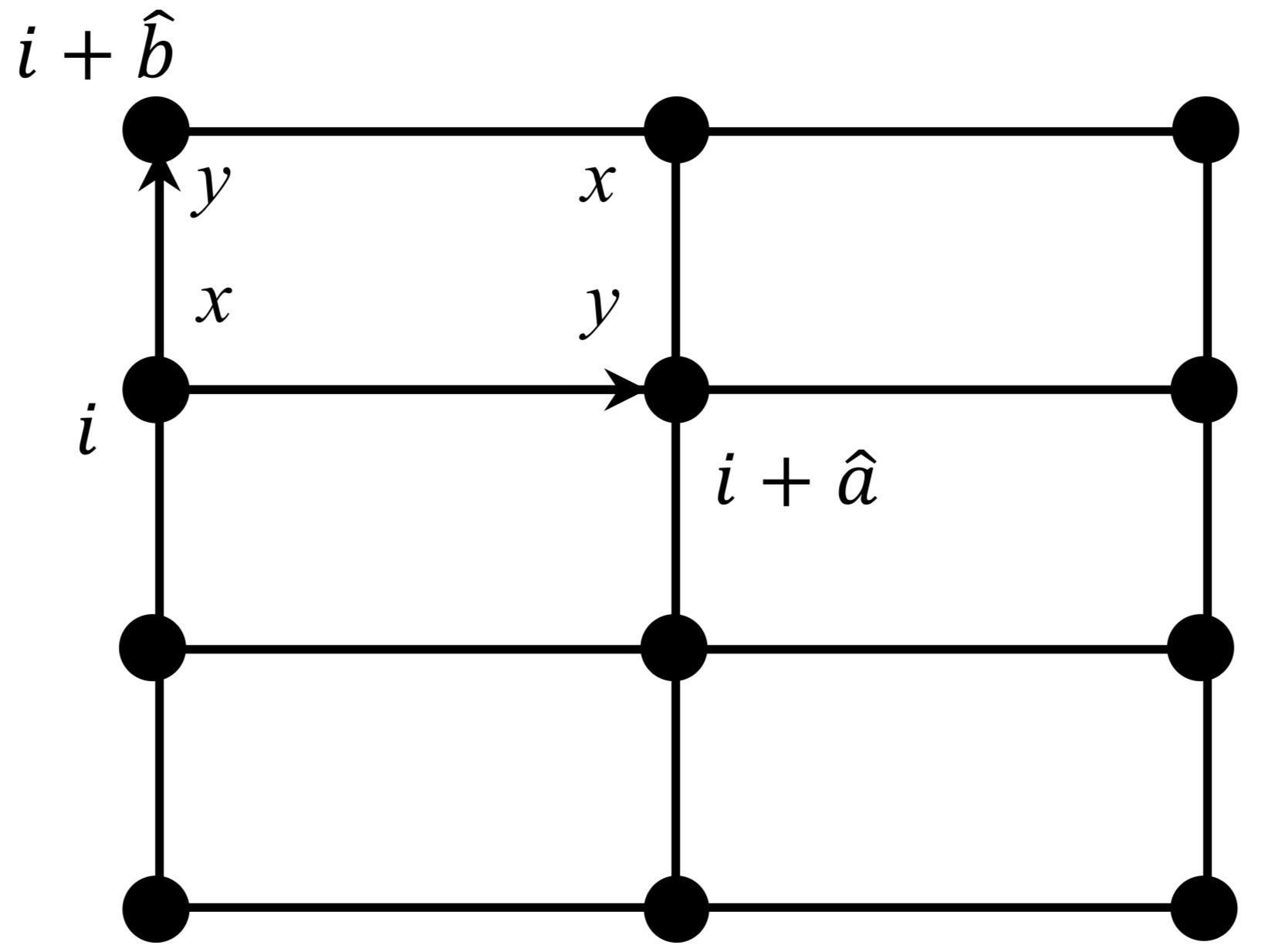}
              \end{center}
  \vskip -0.5cm \protect\caption[]
{(color online) (Left)  Brickwall ladders with coupling parameters $|J_1|,|J_2| \gg |\delta J_1|, |\delta J_2|, |J_3|$; (Right) Wen's toric code manifested in effective spin space. }
    \label{fig:toric}
    \vskip -0.5cm
\end{figure}

Here we show how to implement Wen's two-dimensional toric code \cite{Wen} with our coupled-box clusters.
In Fig.~\ref{fig:3box} (a) if we set $J_4 = J_{4'} = 0$, only one term remains in the perturbation (\ref{eq:p}):
\begin{gather}
{\cal H}_{\text{eff}}^{(4)} = g  \left< \sigma_{1}^z \sigma_{3}^z \sigma_{1'}^z \sigma_{3'}^z \sigma_{2}^y \sigma_{1'}^y \sigma_{4}^x \sigma_{3'}^x\right>_{\text{eff}} = g \hat{F},
\end{gather}
with $g = - \delta J_1 \delta J_2 J_3^2/[2(|J_1|+|J_2|)^3] < 0$. {Meanwhile, as $J_4^x$ and $J_{4'}^x$ vanish together local $J_3^x$ noises do not contribute to $H_{\text{eff}}^{(4)}$.} Recalling that $\Upsilon^\dagger$ in Appendix~\ref{sec:pert} maps each strong bond into one effective $1/2$-spin (see Fig.~\ref{fig:3box} (c)): $\left| \alpha \alpha \right>_{x,(1,2)} \to  \left| \alpha \right>_{x,D}$, $\left| \beta \beta \right>_{x,(1',2')} \to  \left| \beta \right>_{x,C}$, $\left| \gamma \gamma \right>_{y,(3,4)} \to  \left| \gamma \right>_{y,A}$, $\left| \delta \delta \right>_{y,(3',4')} \to  \left| \delta \right>_{y,B}$, in a loop of four effective spins we obtain, 
\begin{gather}
\hat{F} = \left< \sigma_1^z \sigma_2^y \sigma_3^z \sigma_4^x \sigma_{1'}^x \sigma_{3'}^y \right>_{\text{eff}}=  \tau_A^x \tau_B^y \tau_C^x \tau_D^y,
\end{gather}
where $\tau^\nu (\nu=x,y,z)$ are spin operators acting on the effective space (see Fig.~\ref{fig:3box} (c)). Based on this minimal cell with zero $J_4$ and $J_{4'}$, we can then build the two-dimensional lattices of coupled brick-wall ladders shown in Fig.~\ref{fig:toric} Left and reach the Hamiltonian of Wen's toric code in Fig.~\ref{fig:toric} Right:
\begin{gather}
{\cal H} = g\sum_i \hat{F}_i, \qquad \hat{F}_i = \tau_i^x\tau_{i + \hat{a}}^y \tau_{i + \hat{a} + \hat{b}}^x \tau_{i+\hat{b}}^y,
\end{gather}
where $i = (i_a, i_b)$ denotes the square lattice sites. As each $\hat{F}_i$ commutes with each other, it is an exactly solvable model with the ground state configuration $F_i = +1,  \forall i$ for $g < 0$.

\label{sec:flux}

The excitations could be engineered in two ways. On one hand, in the effective spin space the local magnetic field $\sigma_i^x$ or $\sigma_i^y$ acting on the strong $x$ or $y$ bond {(which could be achieved by an inductive or capacitive coupling to a small DC constant bias voltage as before)} becomes the local operation $\hat{X}$ or $\hat{Y}$ which flips the spin on a single site. It creates a diagonal pair of excitations with two corresponding loop-qubit states changing from $+1$ to $-1$. On the other hand, picking up a {single} vertical bond {labelled as} $J_{3'}$ and changing its sign to $-J_{3'}$ {via $\Phi_{3'}$} could introduce a neighboring pair of excitations ({during the process the non-zero $X$} coupling on this isolated vertical bond remains irrelevant). One can also relate Wen's toric code to Kitaev's toric code by moving spins from square lattice sites to the edges of a dual square lattice and performing unitary rotations.

\subsection{SYK {loop} model and Random Ising models}
\label{sec:syk}

For the original SYK model with quenched disorder, the Hamiltonian has the form:
\begin{gather}
	{\cal H} = \frac{1}{4!} \sum_{i,j,k,l = 1}^N J_{ijkl} d_i d_j d_k d_l,
\end{gather}
where the couplings obey Gaussian distribution $P\left( J_{ijkl}\right) \sim \exp \left( -N^3J^2_{ijkl}/12J^2\right): \overline{J^2_{ijkl}} = 3!J^2/N^3, \overline{J_{ijkl}} = 0$. The SYK model is found to be maximally chaotic and share the same Lyapunov exponent of a black hole in Einstein gravity \cite{KitaevnewJosephine}.

\begin{figure}[t]
  \begin{center}
    \includegraphics[width=8cm]{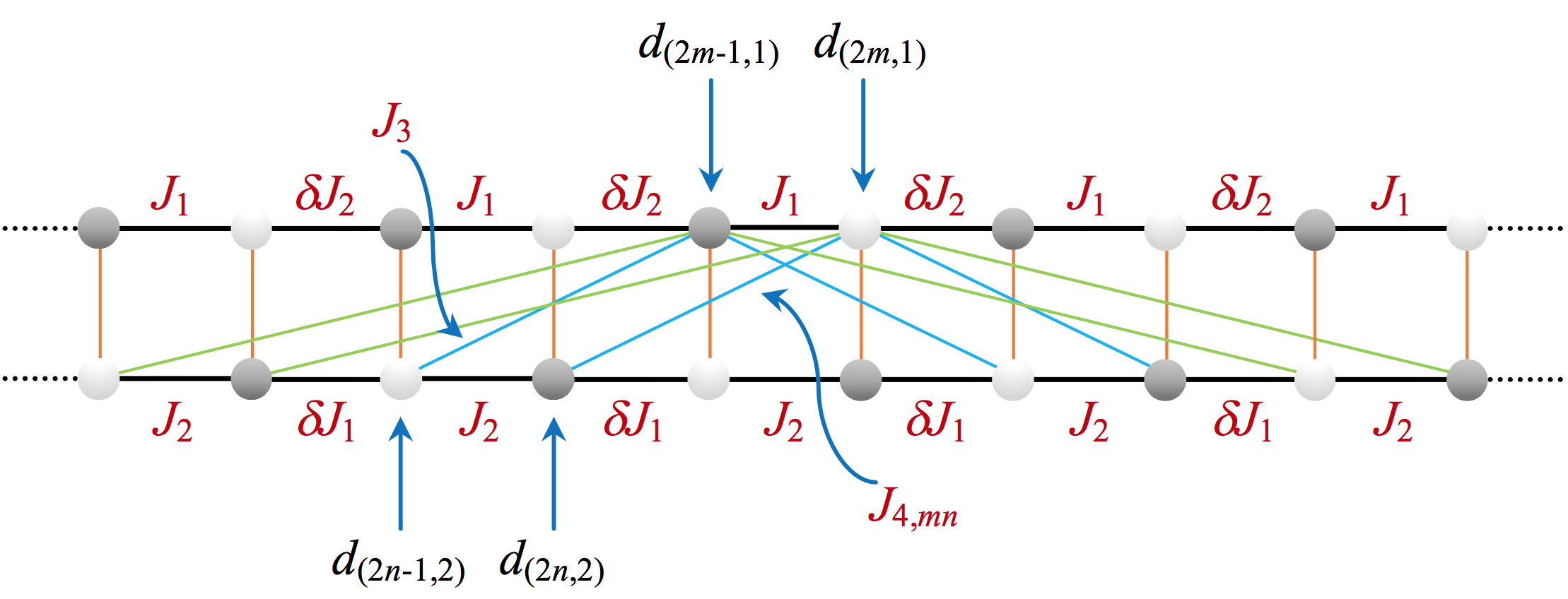}\\
        \includegraphics[width=6.55cm]{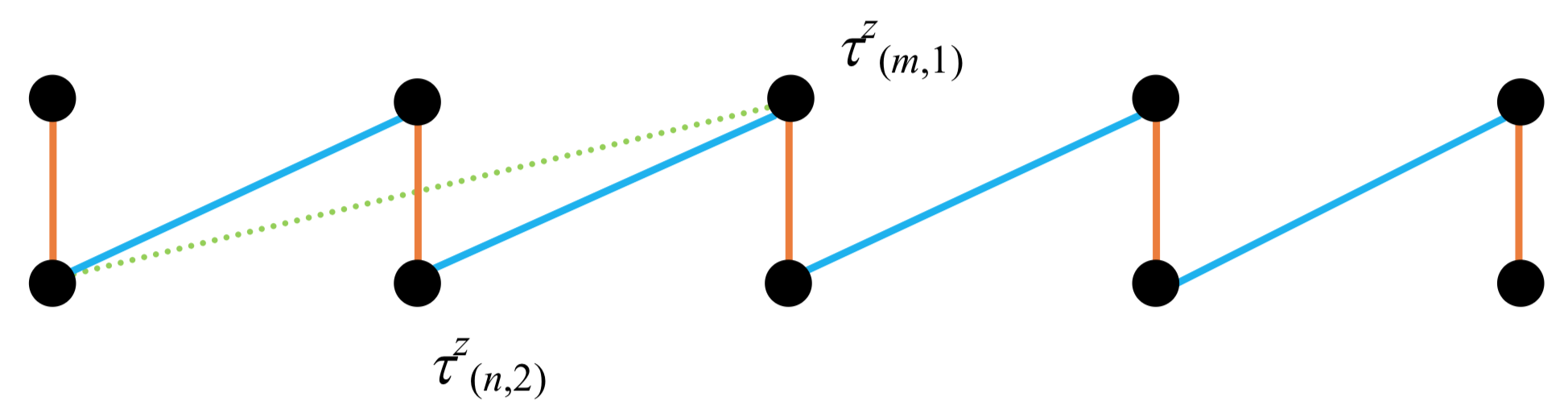}
                  \end{center}
  \vskip -0.5cm \protect\caption[]
  {(color online) (Top) Proposal to approximate the SYK model. The blue and green boxes describe longer-range couplings; (Bottom) Mapping to the long-ranged Ising model.}
    \label{fig:syk}
    \vskip -0.5cm
\end{figure}

By coupling two chains with strong $x$-links and $y$-links by  weak $z$-links shown in Fig.~\ref{fig:syk}, we find two interesting limits to build up the effective Hamiltonian. We define $x = (|J_1| + |J_2|)^{-1} $ as a small number and therefore quantify the weak couplings through: $ \{|J_3|, |J_4| \} = \mathcal{O}(x^s),  \{|\delta J_1|, |\delta J_2| \} = \mathcal{O}(x^t), s,t \in \mathbb{N}^+$.

When $ s \le t$, we can restrict the system to the second-order perturbation in Eq.~(\ref{eq:ho}) and reach an effective Hamiltonian {$\mathcal{O}(x^{2s+1})$}:
\begin{gather}
 {\cal H}^{(2)}_{\text{eff}} =  \sum_{m,n=1}^N J_{mn} d_{(2m-1, 1)}d_{(2m, 1)}d_{(2n-1, 2)}d_{(2n, 2)}, \label{eq:hsyk}
\end{gather}
where the subscript $(j, \alpha)$ denotes the site on the $j$-th column of chain $\alpha = 1,2$ and $J_{mn} = -{J_3J_{4, mn}}/{(|J_1| + |J_2|)}$. The coupling constants {$J_{mn}$} are random variables with a Gaussian distribution ensured by the adjustability of 
{$\Phi_{4, mn}$}: $P\left( J_{mn}\right) \sim \exp \left( -NJ^2_{mn}/2J^2\right)$. $\left[id_{(2m-1,\alpha)}d_{(2m,\alpha)}, H_{\text{eff}}\right] = 0$ and $(id_{(2m-1,\alpha)}d_{(2m,\alpha)})^2 = 1$ imply that $id_{(2m-1,\alpha)}d_{(2m,\alpha)}$ is a good quantum number with the value $\pm 1$. We arrive at the following map:
\begin{gather}
	{\cal H}^{(2)}_{\text{eff}} = \sum_{m,n=1}^N J_{mn} \tau^z_{(m,1)}\tau^z_{(n,2)},
\end{gather}
where $\tau_{(m,\alpha)}^z = id_{(2m-1,\alpha)}d_{(2m,\alpha)}$. This gives rise to a one-dimensional Ising model (e.g. the zigzag path formed by orange {loops} and half of blue loops shown in Fig.~\ref{fig:syk} Bottom) with long-range {random} interactions (for example, green loops).
Following the mapping to effective spin space as in Sec.~\ref{sec:wen}, we can get the same result and take into account higher order corrections. Back to two coupled boxes in Fig.~\ref{fig:3box} (a), {from Eqs.~(\ref{eq:ho}) and (\ref{eq:p})} we find
$ \mathcal{P}_1 = \langle \sigma_1^z\sigma_2^z\sigma_3^z\sigma_4^z\rangle_{\text{eff}} = \tau_D^z\tau_A^z,
 \mathcal{P}_3 = \langle \sigma_{1'}^z\sigma_{2'}^z\sigma_{3'}^z\sigma_{4'}^z\rangle_{\text{eff}} = \tau_C^z\tau_B^z$,
which recovers the classical Ising couplings shown in Fig.~\ref{fig:3box} (b). Quantum corrections arise from the fourth-order perturbation with the {terms}:
$\mathcal{P}_1\mathcal{P}_3 =\tau_A^z\tau_B^z\tau_C^z\tau_D^z, 
\mathcal{P}_{\widetilde{123}} =\tau_A^x\tau_B^x\tau_C^y\tau_D^y, 
\mathcal{P}_{2} =\tau_A^y\tau_B^y\tau_C^x\tau_D^x,
\mathcal{P}_{\widetilde{12}} =\tau_A^x\tau_B^y\tau_C^x\tau_D^y, 
\mathcal{P}_{\widetilde{23}} =\tau_A^y\tau_B^x\tau_C^y\tau_D^x$.
 {Noises from non-zero $X$ couplings on vertical bonds} would produce a small magnetic field along $z$ direction on sites $A$ and $B$, as the effective interactions $\langle \sigma_3^x \sigma_4^x \rangle \sim \tau_A^z, \langle \sigma_{3'}^x \sigma_{4'}^x \rangle \sim \tau_B^z$.

When $s>t$, we can drop out the terms $\sim {\mathcal{O}(x^{4s+3})}$ in the fourth-order perturbation of Eq.~(\ref{eq:ho}) and the effective Hamiltonian has the form {$\mathcal{O}(x^{2s+2t+3})$}:
\begin{gather}
	 {\cal H}^{(4)}_{\text{eff}} = \sum_{m,n=1}^N  \sum_{l=1}^4 J_{mnl} \mathcal{P}_{d}^{mnl},
\end{gather}
with coefficients $J_{mn1}= - \frac{\delta J_1 \delta J_2 J_3^2}{2(|J_1|+|J_2|)^3}$, $J_{mn2}= - \frac{5\delta J_1 \delta J_2  J_3 J_{4,(m+1)n}}{2(|J_1|+|J_2|)^3}$,
	$J_{mn3}= - \frac{\delta J_1 \delta J_2 J_3 J_{4,mn}}{2(|J_1|+|J_2|)^3}$,  $J_{mn4}= - \frac{\delta J_1 \delta J_2 J_{4,mn}J_{4,(m+1)n}}{2(|J_1|+|J_2|)^3}$. Here $\mathcal{P}_{d}^{mnl}$ is the loop operator which denotes the $4$-body couplings between $d$-Majoranas living on the vertices of ``tilted'' boxes: $ \mathcal{P}_{d}^{mnl} = d_{(2m-1, 1)}d_{(2m+l, 1)}d_{(2n-1, 2)}d_{(2n+l, 2)}  (l = 1,2)$,
	  $\mathcal{P}_{d}^{mn3} = d_{(2m, 1)}d_{(2m+1, 1)}d_{(2n, 2)}d_{(2n+1, 2)}$, $\mathcal{P}_{d}^{mn4} = d_{(2m, 1)}d_{(2m+2, 1)}d_{(2n, 2)}d_{(2n+2, 2)}$. 
This model could reveal glassy phases of the Ising model and quantum corrections could be controlled through effective fourth-order corrections, which will be studied in a future work. An analogue of the Anderson-Edwards \cite{AE} order parameter could be measured as well as echo spin measurements \cite{echo}. Links with many-body localization phenomena could also occur \cite{Pollmann}.

\section{Conclusion}
To summarize, we suggest a superconducting toolbox starting from spin degrees of freedom (qubits) to study the formation of ${\cal Z}_2$ quantum spin liquids and many-body Majorana states.  Spin correlations can be measured with current technology \cite{Roushan,Neill} and local susceptibility measurement through the NMR device could reveal the occurrence of Majorana degrees of freedom and quantum phase transitions. We have addressed detuning and dissipation effects and observed that the emergent gauge fields could be detected on several Floquet periods, even though the quantization of the fluxes could be altered.  We have discussed the protection of the different phases related to possible detuning effects. In lattices of several boxes, quantum spin liquid states are associated with a N\' eel order of gauge fields making analogies with Ising models. These Ising models can be disordered by engineering local fluxes and one could realize various glassy phases in relation with the SYK Majorana model.  As other practical applications, we have built relations with the Wen's toric code in brickwall ladders. This box at a boundary could allow us to study other quantum impurity Majorana models by analogy {with} Kondo models (with four spins S=1/2 or two spins S=1). We also note another proposal to engineer four-body Ising interactions with Josephson junctions \cite{Puri}. It is also promising to see that the occurrence of orbital loop currents in Mott insulators \cite{Philippe,cuprates} {has} now been observed. Realizing anistropic spin coupling constants in two dimensions is also possible in cold atoms \cite{Duan,Juliette}. 

\bigskip

Acknowledgements: This work has benefitted from useful discussions with D. Bernard, J. Esteve, J. Gabelli, T. Goren, L. Herviou, S. Munier, C. Mora, C. Neill, A. Petrescu, O. Petrova, K. Plekhanov, P. Roushan, at the DFG meeting FOR2414 in Hamburg and at the Conference in Milton Keynes England on topological photon systems. Supports by the Deutsche Forschungsgemeinschaft via DFG FOR 2414 and by the LABEX PALM Grant No. ANR-10-LABX-0039 are acknowledged. LH acknowledges support from the Ministry of Economy and Competitiveness of Spain through the ``Severo Ochoa'' program for Centres of Excellence in R\&D (SEV-2015- 0522), Fundaci\'{o} Privada Cellex, Fundaci\'{o} Privada Mir-Puig, and Generalitat de Catalunya through the CERCA program.

\appendix

\section{Table of Parameters}
\label{app:para}
\begin{table}[h]
\begin{center}
\caption{Parameters for box circuit}  \label{tab:para}
\begin{tabular}{cc|cc}
\hline
\hline
Parameter & Relation & Parameter & Relation  \\
\hline
$\lambda$ & $\left( E_{J_q}/(2E_{C_q}) \right)^{1/4}$ & $J_1$ & $-2E_L/(\lambda_A \lambda_B)$\\
$s$ & $C/C_q$ & $J_2$ & $-2E_Cs_As_B\lambda_A\lambda_B$ \\
$\omega_{q}$ & $\sqrt{8E_{C_{q}}E_{J_{q}}}/ \hbar$ &  $J_3$ & $-E_{J,3}/(2\lambda_A\lambda_B)^2$ \\
$\omega_{L'}$ & $ 4E_{L'} V_{0} / (\hbar\phi_0\lambda\omega)$ & $J_4$ & $-E_{J,4}/(2\lambda_A\lambda_B)^2$ \\
$\omega_{C'}$ & $2V_{0}es'\lambda/\hbar$ &   $J_3^x$ & $-E_{J,3}/(\lambda_A\lambda_B)$ \\
$\epsilon_q$ & $\left(\hbar \omega_q - E_{C_q}\right)/2$ & $J_4^x$ & $-E_{J,4}/(\lambda_A\lambda_B)$\\
$\epsilon_{L}$ & $E_{L}/\lambda^2$ & $\tilde{J}_3^x$ & $-2E_{\tilde{L},3}/(\lambda_A\lambda_B)$  \\
$\epsilon_{C}$ & $E_{C}(s \lambda)^2$ & $\tilde{J}_4^x$ &  $-2E_{\tilde{L},4}/(\lambda_A\lambda_B)$  \\
$\epsilon_J$ & $-E_{J,\bot}/(2\lambda^2)$  & $J'_{\nu}$ & $r_{\nu}J_{\nu}$\\
       \hline
\hline
\end{tabular}  \\
\smallskip
\footnotesize{* Notation of subscripts: $A$ for sites $\{1, 4\}$, $B$ for sites $\{2, 3\}$, $\nu = x, y, z$.}
\end{center}
\label{tab:bc}
\end{table}
Our dynamical protocols simulated in numerics are designed to study spin observables and detect $\mathcal{Z}_2$ gauge fields. It is important to analyze the constraints in terms of experimental parameters. For simplicity, here we suppress the site index $j$. From Table~{\ref{tab:para}}, the limit of weak vertical bonds $|J_1|, J_2| \gg |J_3|, |J_4|$ requires $\lambda \gg 1 \gg s, s\lambda^2 \sim 1, E_L, E_C \gg E_{J, 3}, E_{J, 4}$.
The main contribution to the magnetic field $\sigma^z$ comes from the transition frequency of the qubit $\hbar\omega_q \gg E_L, E_C, E_J, E_{\tilde{L}}, E_{L'}, E_{C'}$. 
To cancel this local field, we engineer a circularly polarized field and impose $\omega_1 = \omega_{L'} = \omega_{C'}$ giving rise to $4E_{L'} = s'\lambda^2 \hbar \omega$
with $\hbar\omega \gg E_{L'}, \lambda \gg 1 \gg s',  1 \gg s'\lambda^2$. 

We further choose a particular combination of frequencies from Eq.~{(\ref{eq:wcondition})}: $\omega_1 = \sqrt{2} \omega_0$, $\omega = 3\omega_0/2$. It results in $V_0 = {3\sqrt{2}\phi_0 \hbar \omega_0^2 \lambda}/{(8E_{L'})}$. Since  $\omega_0 \gg E_{L'}, \lambda \gg 1$, both the amplitude $V_0$ and frequency $\omega$ of the AC driving device should be large. 
Additionally, it is also noted that inside the NMR, the plasma frequency $\omega_{P}$ is much smaller compared to $\omega$:
$\omega_P \sim {1}/{\sqrt{L'C'}} \sim \sqrt{{E_{L'}}/{C'}} \ll \omega \sim \omega_0 \sim \omega_q \sim \sqrt{{E_{J_q}}/{C_q}}$,
which leads to ${E_{L'}}/{E_{J_q}} \ll s' \ll 1$. It is consistent with our limit of large $\lambda \gg 1$.

\section{NMR Unitary Transformation}
\label{sec:om}
Here we present some useful mathematical formulas related to the gauge transformation in Sec.~\ref{sec:nmr}. Spin operators commute on different sites, so do $F_j(t)$. It enables us to suppress site indices $j$ and focus on the single spin problem:
\begin{gather}
{\cal H}_C(\tau)= \omega_{0} S_z  - \omega_{1} \left(  \cos \tau  S_x + \sin \tau  S_y \right), \notag \\
G_C = e^{iF}{\cal H}_Ce^{-iF} + i\hbar \omega \left( \partial_\tau e^{iF}\right) e^{-iF}.  \label{eq:gc}
\end{gather}
Applying the Baker-Campbell-Hausdorff formula,
\begin{gather}
\begin{split}
e^{iF}{\cal H}_Ce^{-iF} &= {\cal H}_C + i\left[ F, {\cal H}_C\right] + \frac{i^2}{2!} \left[ F, \left[ F,{\cal H}_C\right] \right]  \\
  &\phantom{=} + \frac{i^3}{3!} \left[ F,  \left[ F, \left[ F, {\cal H}_C\right] \right] \right] + \cdots,  \\
\left( \partial_\tau e^{iF}\right) e^{-iF} &= \partial_\tau \left( \sum_{n=0}^{\infty} \frac{(iF)^n}{n!}\right)  e^{-iF} \\
&= i\partial_\tau F + \frac{i^2}{2!} \left[F, \partial_\tau F \right]+ \frac{i^3}{3!}\left[ F, \left[F, \partial_\tau F \right] \right] + \cdots. \label{eq:ee}
\end{split}
\end{gather}
Now we assume $F(\tau)$ is a linear function of $S_i \ (i = x,y,z)$ as $H_C(\tau)$:
\begin{gather}
F(\tau) = l(\tau)  S_x + m(\tau)  S_y + n(\tau)  S_z. \label{eq:ftau}
\end{gather}
Due to the closed $\mathfrak{su}(2)$ algebra for spin-1/2
\begin{gather}
\left[ S_i, S_j \right] = i\hbar \epsilon_{ijk}S_k,
\end{gather}
$G_C$ is also linear in $S_i$. For an arbitrary linear function $Q \left(S_i\right)$, we find
\begin{gather}
\left[ F,  \left[ F, \left[ F, Q\right] \right] \right] = \alpha^2 \left[ F, Q \right], \quad \alpha^2 = \hbar^2 \left( l^2 + m^2 + n^2 \right).
\end{gather}
Then the infinite series in $G_C$ can be grouped into the finite expression:
\begin{gather}
\begin{split}
&G_C = {\cal H}_C + \frac{\sin \alpha}{\alpha} i \left[ F, {\cal H}_C\right] + \frac{\cos \alpha -1}{\alpha^2} \left[ F, \left[ F, {\cal H}_C\right] \right]  \\
&+ \hbar \omega \left(  -\partial_\tau F + \frac{\cos \alpha -1}{\alpha^2}i \left[ F, \partial_\tau F\right]  - \frac{\sin \alpha - \alpha}{\alpha^3}\left[ F, \left[F, \partial_\tau F \right] \right] \right). 
\end{split}\label{eq:gcexact}
\end{gather} 
Taking $F(\tau) = \alpha \left(\sin (\tau)  S_x - \cos (\tau)  S_y\right)/\hbar$, we derive the expression of $G_C$ in Eq.~(\ref{eq:gco}). In the same manner, a single local spin operator $S_i$ is transformed into the rotating frame through
\begin{gather}
e^{iF}S_ie^{-iF} = S_i + \frac{\sin \alpha}{\alpha} i \left[ F, S_i\right] + \frac{\cos \alpha -1}{\alpha^2} \left[ F, \left[ F, S_i\right] \right]. \label{eq:si}
\end{gather}

\section{Perturbation Theory Study}

\label{sec:pert}
In perturbation theory, a system of two coupled boxes in Fig.~\ref{fig:3box} (a) consists of the interaction terms:
\begin{gather}
{\cal H}_0 = J_1\left( \sigma_1^x \sigma_2^x + \sigma_{1'}^x \sigma_{2'}^x\right) + J_2 \left( \sigma_3^x \sigma_4^x + \sigma_{3'}^x \sigma_{4'}^x\right), \notag \\
V = {\cal \delta H}_{\bot} + {\cal \delta H}_{\parallel}, \notag \\
{\cal \delta H}_{\bot} = J_3 \left( \sigma_1^z \sigma_3^z + \sigma_{1'}^z \sigma_{3'}^z \right) +  \left( J_4  \sigma_2^z\sigma_4^z + J_{4'} \sigma_{2'}^z\sigma_{4'}^z \right), \notag \\
{\cal  \delta H}_{\parallel} =  \delta J_2 \sigma_2^y \sigma_{1'}^y + \delta J_1 \sigma_4^x \sigma_{3'}^x.
 \end{gather}
Here $(J_1, J_2) \ll -1, (\delta J_1, \delta J_2, J_3, J_3^x) \to ( 0^{-}, 0^{-}, 0^{+}, 0)$ and $J_4, J_{4'}$ can be controlled around $0^{\pm}$  by the phases $\Phi_4, \Phi_{4'}$. We notice in Sec.~\ref{sec:per} when suppressing the vertical $X$ couplings on $J_3$ bonds, $\left( J^x_4 \sigma_2^x\sigma_4^x +  J^x_{4'}\sigma_{2'}^x\sigma_{4'}^x \right)$ become irrelevant operators in any order of perturbation, thus we have ignored them in $\delta \mathcal{H}_{\bot}$.

The ground state of $\mathcal{H}_0$ is constructed by four effective spins: $\left|\alpha \alpha \right>_{x,(1,2)} \otimes \left|\beta \beta \right>_{y,(3,4)} \otimes \left|\gamma \gamma \right>_{x,(1',2')} \otimes \left|\delta \delta \right>_{y,(3',4')}$ ($\alpha, \beta, \gamma, \delta = \pm 1$). We introduce a map $\Upsilon$: $\Upsilon \left| \alpha \right> =  \left| \alpha \alpha \right> $ and find ${\cal H}_{\text{eff}}^{(0)} = 2\left(J_1 + J_2\right)$,  ${\cal H}_{\text{eff}}^{(1)} = \Upsilon^\dagger V \Upsilon = 0$, ${\cal H}_{\text{eff}}^{(3)}  = \Upsilon^\dagger V G_0' V G_0' V \Upsilon = 0$ where $G_0'(E) = \left((E- {\cal H}_0)^{-1}\right)'$. The non-zero contributions arise from the second and fourth orders
\begin{gather}
\begin{split}
{\cal H}_{\text{eff}}^{(2)}  = &\Upsilon^\dagger V G_0' V \Upsilon = \text{cst} - \frac{J_3J_4}{|J_1| + |J_2|}  \left< \sigma_1^z \sigma_2^z \sigma_3^z \sigma_4^z \right>_{\text{eff}} \\
&- \frac{J_3J_{4'}}{|J_1| + |J_2|}\left< \sigma_{1'}^z \sigma_{2'}^z \sigma_{3'}^z \sigma_{4'}^z \right>_{\text{eff}} ,\\
{\cal H}_{\text{eff}}^{(4)} = &\Upsilon^\dagger V G_0' V  G_0' V G_0' V \Upsilon  \\
= &\text{cst} - \frac{1}{2(|J_1|+|J_2|)^3} \left( J_3J_4(J_3^2+ J_{4'}^2) \left< \sigma_1^z \sigma_2^z \sigma_3^z \sigma_4^z \right>_{\text{eff}} \right. \\
&+ 2J^2_3J_4J_{4'} \left< \sigma_1^z \sigma_3^z \sigma_2^z \sigma_4^z \sigma_{1'}^z \sigma_{3'}^z \sigma_{2'}^z \sigma_{4'}^z \right>_{\text{eff}} \\
&\left. + J_3J_{4'}(J_3^2+ J_{4}^2) \left< \sigma_{1'}^z \sigma_{2'}^z \sigma_{3'}^z \sigma_{4'}^z \right>_{\text{eff}}\right)  \\
&- \frac{\delta J_1 \delta J_2 }{2(|J_1|+|J_2|)^3} \left( 5J_3 J_{4'} \left< \sigma_1^z \sigma_3^z \sigma_{2'}^z \sigma_{4'}^z \sigma_{2}^y \sigma_{1'}^y \sigma_{4}^x \sigma_{3'}^x  \right>_{\text{eff}} \right. \\
&+ J_3J_4\left< \sigma_{2}^z \sigma_{4}^z \sigma_{1'}^z \sigma_{3'}^z \sigma_{2}^y \sigma_{1'}^y \sigma_{4}^x \sigma_{3'}^x \right>_{\text{eff}}  \\
& + J_3^2\left< \sigma_{1}^z \sigma_{3}^z \sigma_{1'}^z \sigma_{3'}^z \sigma_{2}^y \sigma_{1'}^y \sigma_{4}^x \sigma_{3'}^x\right>_{\text{eff}}  \\
&\left.+ J_4 J_{4'} \left< \sigma_{2}^z \sigma_{4}^z \sigma_{2'}^z \sigma_{4'}^z \sigma_{2}^y \sigma_{1'}^y \sigma_{4}^x \sigma_{3'}^x\right>_{\text{eff}}\right). \label{eq:p}
\end{split}
\end{gather}

\clearpage

\end{document}